\documentclass[traditabstract]{aa}
\usepackage{graphicx}
\usepackage{lscape}      
\usepackage{txfonts}     
\usepackage{longtable}   
\usepackage{color}
\usepackage{natbib}
\usepackage{ulem}
\bibpunct{(}{)}{;}{a}{}{,} 

\def\cone {\ifmmode{{\rm C}{\rm \small I}(^3\!P_1\!-^3\!P_0)}
     \else{C\ts {\scriptsize I}{\small$(^3\!P_1\!-^3\!\!\!P_0)$}}\fi}
\def\ctwo {\ifmmode{{\rm C}{\rm \small I}(^3\!P_2\!-^3\!P_1)}
     \else{C\ts {\scriptsize I}{\small$(^3\!P_2\!-^3\!\!\!P_1)$}}\fi}

\def\tex {\ifmmode{{T}_{\rm ex}}\else{$T_{\rm ex}$}\fi}
\def\tmb {\ifmmode{{T}_{\rm mb}}\else{$T_{\rm mb}$}\fi}
\def\ci     {\ifmmode{{\rm C}{\rm \small I}}\else{C\ts {\scriptsize I}}\fi}
\def\hi     {\ifmmode{{\rm H}{\rm \small I}}\else{H\ts {\scriptsize I}}\fi}
\def\hh     {\ifmmode{{\rm H}_2}\else{H$_2$}\fi}

\def\ts     {\thinspace}
\def\kms    {\ifmmode{{\rm \ts km\ts s}^{-1}}\else{\ts km\ts s$^{-1}$}\fi}
\def\msol   {\ifmmode{{\rm M}_{\odot}}\else{M$_{\odot}$}\fi}
\def\lsol   {\ifmmode{{\rm L}_{\odot}}\else{L$_{\odot}$}\fi}
\def\zsol   {\ifmmode{{\rm Z}_{\odot}}\else{Z$_{\odot}$}\fi}

\begin{document}

\title{Debris disks around M dwarfs: The \textit{Herschel} DEBRIS survey}

\subtitle{}

\author{Jean-Fran\c cois Lestrade  \inst{1}
\and
B.C. Matthews \inst{2}
\and 
G.M. Kennedy \inst{3}
\and
B. Sibthorpe  \inst{4,5}
\and
M.C. Wyatt  \inst{6}
\and
M. Booth  \inst{4,7}
\and
J.S. Greaves \inst{8}
\and
G. Duch\^ene \inst{9,10}
\and 
A. Moro-Mart\'in  \inst{11,12}
\and
C. Jobic   \inst{13,1}
}
%
\institute{
LUX, Observatoire de Paris, Universit\'e PSL, Sorbonne Universit\'e,
CNRS, 75014 Paris, France
\and
Herzberg Institute of Astrophysics (HIA), National Research Council of Canada, Victoria, BC, Canada               
\and
Department of Physics and Centre for Exoplanets and Habitability, University of Warwick, Gibbet Hill Road, Coventry CV4 7AL, UK
\and
UK Astronomy Technology Centre (UKATC), Royal Observatory Edinburgh, Blackford Hill, Edinburgh, EH9 3HJ, UK
\and
Airbus Defence and Space, Gunnels Wood Rd, Stevenage SG1 2AS, UK
\and
Institute of Astronomy (IoA), University of Cambridge, Madingley Road, Cambridge, CB3 0HA, UK
\and
Astrophysikalisches Institut und Universit\"atssternwarte, Friedrich-Schiller-Universit\"at Jena, Schillerg\"a\ss chen 2-3, D-07745 Jena, Germany
\and
School of Physics \& Astronomy, CHART, Cardiff University, Cardiff, UK
\and
Univ. Grenoble Alpes, CNRS, IPAG, 38000 Grenoble, France.
\and
Astronomy Department, UC Berkeley, 601 Campbell Hall, Berkeley CA 94720-3411, USA
\and
Space Telescope Science Institute, 3700 San Martin Dr., Baltimore, MD 21218, USA
\and
Physics and Astronomy, Center for Astrophysical Sciences, Johns Hopkins University, Baltimore, MD 21218, USA
\and
CNES, 18 avenue Edouard Belin, 31400 Toulouse, France
}
\offprints{J-F Lestrade, \email{jean-francois.lestrade@obspm.fr}}

\date{Received 27/07/2024 ; accepted 16/12/2024}
\titlerunning{Debris disks around M-dwarfs : Herschel}

\abstract{The {\it Herschel} open-time key program  Disc Emission via a Bias-free Reconnaissance in the
Infrared and Sub-millimeter (DEBRIS)
is an unbiased survey of the nearest $\sim$ 100 stars for each stellar type A-M
observed with a uniform  photometric sensitivity to search for cold debris disks around them.
The analysis of the Photoconductor Array Camera and Spectrometer (PACS) photometric observations 
of the 94 DEBRIS M dwarfs of this program 
is presented in this paper, following upon two companion papers on the DEBRIS A-star and FGK-star subsamples.
In the M-dwarf subsample, two debris disks have been detected, around  the M3V dwarf 
GJ~581 and the M4V dwarf Fomalhaut~C (LP~876-10). This result gives a disk detection rate of $2.1^{+2.7}_{-0.7}$~\% at the 
68\% confidence level, significantly less than measured for earlier stellar types in the  DEBRIS program.
However, we show that the survey of the DEBRIS M-dwarf subsample 
is about ten times shallower than the surveys of the DEBRIS FGK subsamples  
when studied in the physical parameter space of the disk's fractional dust luminosity  
versus blackbody radius.
Furthermore, had the DEBRIS K-star subsample been observed at the same shallower depth in this parameter space,
its measured disk detection rate would have been statistically consistent with the one 
found for the M-dwarf subsample. Hence,  the incidence of debris disks 
does not appear to drop  from the K subsample to the M subsample of the DEBRIS program, when considering disks in the same region 
of physical parameter space. An alternative explanation is that the only two bright disks discovered in the M-dwarf subsample
would not, in fact,  be statistically representative of the whole population.
}


\titlerunning{Debris disks around M-type stars}
\maketitle

\section{Introduction} \label{intro}

A debris disk is a key component of a planetary system and 
a tracer of the history of its formation and its dynamics. It is made up of
the surplus of km-sized planetesimals that were not able to agglomerate into larger planetary mass bodies during 
the protoplanetary phase.
The Solar System's Asteroid belt and Edgeworth-Kuiper belt are debris disks.
The remnant planetesimals are not directly detectable when surrounding stars other than the Sun, however  
their orbit crossings at high velocity can result in destructive collisions that  
continuously regenerate micrometer-sized dust. This dust may reflect or absorb and reemit enough star light to
make the disk  observable. The physical and observational  properties of debris disks  
have been reviewed  in \citet{Lagr00}, \citet{Wyat08}, \citet{Kriv10}, \citet{Matt14}, \citet{Hugh18}, and \citet{Wyat20}.
  
Surveys in the mid-infrared and far-infrared domains with {\it IRAS}  \citep[e.g.,][]{Oudm92,Mann98}
and {\it Spitzer} \citep[e.g.,][]{Tril08,Su06} have shown that debris disks can be found 
 around main sequence stars of all stellar types. Given these stars exhibit differences in stellar luminosities on the level of several orders of magnitude implies that planetesimal formation 
---  a critical step in planet formation ---  is a robust process that can take place under a wide range of conditions.
However, it is also widely known that debris disks have been more seldomly detected around M-type dwarfs than 
earlier type stars. This was previously shown in the surveys 
of M dwarfs in the mid-infrared \citep{Plav05,Plav09}, 
in the far-infrared with {\it Spitzer} \citep{Gaut07}, and at (sub)millimeter wavelengths \citep{Lest06,Lest09}. 
This paucity of observed debris disks around M dwarfs seems surprising since in the early stage of their evolution,
stars of all stellar types have similar incidence of protoplanetary disks according to observations of star forming regions, 
such as Taurus-Auriga and $\rho$~Oph  \citep[e.g.,][]{Andr05}. 

Nonetheless, these pre-{\it Herschel} surveys of debris disks have been biased through selections in age, 
metallicity of host stars, and binarity. 
The {\it Herschel} open-time key program  Disc Emission via a Bias-free Reconnaissance in the 
Infrared and Sub-millimeter (DEBRIS) 
was designed to estimate and compare the true incidences 
of debris disks across all stellar types in unbiased subsamples; 
the nearest $\sim$100 stars of each stellar type A-F-G-K-M have been surveyed to search for possible surrounding debris disks \citep{Matt10}
in a flux-limited sample, with a sensitivity that is approximately uniform  across all stellar types in
 the far-infrared domain. This has offered the maximum sensitivity
to cold dust, expected to be between 10 K and 60 K for debris disks with blackbody radii between a few~au 
and a few hundreds of ~au.

This paper presents the analysis of the 94 M dwarfs of the DEBRIS subsample to complement
the analyses of the DEBRIS A-star subsample  in \citet{Thur14} and the DEBRIS FGK-star subsamples in \citet{Sibt18}.
In Sect.\,\ref{obs}, we describe the M-dwarf subsample, observations, and photometric measurements.  
In Sect.\,\ref{incid}, we determine the disk detection rate in this subsample with the {\it Herschel}/Photoconductor Array Camera and Spectrometer ({\it Herschel}/PACS)
 observations.
In Sect.\,\ref{SPIRE}, we present the  {\it Herschel}/Spectral and Photometric Imaging Receiver ({\it Herschel}/SPIRE) observations of 25 M dwarfs of the subsample. 
In Sect.\,\ref{complete}, we study the completeness of detections in the subsample
in the fractional dust luminosity, $f_d$, versus  blackbody radius, $R_{BB}$, parameter space.
In Sect.\,\ref{constrains}, we show that two power laws are able to constrain the probability distributions of $f_d$
and $R_{BB}$ in the DEBRIS A, F, G, K subsamples. In Sect.\,\ref{Dis}, we place the DEBRIS survey  
into perspective, with the pre-{\it Herschel} searches for debris disks around M dwarfs in the far-infrared (FIR) domain 
and at longer wavelengths. We present our conclusions in Sect.\,\ref{Con}.    

\section{Sample and observations } \label{obs}

\subsection{The DEBRIS {\it Herschel} survey}

The full DEBRIS survey 
is a {\it Herschel} open-time key program designed to observe 446 nearby stars of stellar
types A to M \citep{Matt10}. The DEBRIS sample is drawn from the Unbiased Nearby
Stars catalogue (UNS; \citealt{Phil10}). The DEBRIS target list comprises the nearest
systems (all-sky), subject to a cut in the predicted cirrus confusion  level
towards each system ($> 1.2$~mJy at 100~$\mu$m), with details reported in \citet{Phil10}.
Consequently, the DEBRIS subsample which consists of about 100 stars of each stellar type is volume-limited
and free of bias towards any particular stellar parameters or any prior knowledge of
disc or planetary system. Since these are all field stars, and not generally members of clusters or
associations, the stellar ages are uncorrelated.
The M subsample comprises 94 M-stars and is complete to 8.6 pc excepting stars with high cirrus confusion level as mentioned.
The ages of M-stars are notoriously difficult to establish and are most likely randomly distributed between 
$\sim$100~Myr and $\sim$10~Gyr in the sample.

\subsection{Observations and data reduction}

All targets were observed at 100 and 160~$\mu$m using
the PACS photometer \citep{Pogl10} on board the space 
{\it Herschel} telescope \citep{Pilb10}.
The  {\it mini scan-map} observing mode was used for all PACS observations.
Two scans of each target were performed with a relative
scanning angle of $40^{\circ}$ to mitigate striping artifacts associated with
$1/f$ noise. Scan-maps used a scanning rate of 20 arcsecs per second
and were constructed of 3 arcmins scan legs with a separation of
4 arcsecs between legs. The nominal DEBRIS observations used 8
scan-legs per map and performed two map repeats per scanning direction.
The data were reduced using Version 10.0 of the \textit{Herschel}
Interactive Pipeline Environment (HIPE) 
by \citet{Ott10}. The standard pipeline processing steps were used and maps were made using
the photProject task. The time ordered data were high-pass filtered,
passing scales smaller than 66 arcsecs at  100~$\mu$m and 102
arcsecs at 160~$\mu$m (equivalent to a filter radius of 16 and 25 frames
respectively), to remove low-frequency noise in the scan direction.
Sources of $>$2$\sigma$ were then identified in this first stage map
to create a filter mask. The original data are then filtered a second
time, using the derived mask to exclude bright sources, which would
otherwise result in ringing artifacts, and a final map was then produced.
The mean noise root mean square (rms) values are: 2.0~mJy/6.8$''$beam ($\sigma=0.35$mJy/b on mean) at 100~$\mu$m and 4.3~mJy/11.4$''$beam 
($\sigma=0.98$mJy/b) at 160~$\mu$m.   

In addition, a smaller survey of 25 M dwarfs (the 20 nearest and 5 more of interest, also PACS targets) 
was conducted with SPIRE \citep{Grif10} to search for colder dust in 
larger belts than usually expected. The observations 
used the small-map mode, resulting in simultaneous
250, 350, and 500~$\mu$m images. The data were reduced using
HIPE (version 7.0 build 1931), adopting the natural pixel scale
of 6, 10, 14 arcsecs at 250, 350 and 500~$\mu$m, respectively. Their
noise rms levels are: 6.1~mJy/18.2$''$beam, 7.9~mJy/24.9$''$beam, and
8.3~mJy/36.3$''$beam at 250, 350, and 500~$\mu$m, respectively.

\begin{figure*}[t!]
\resizebox{18cm}{!}{\includegraphics[angle=0] {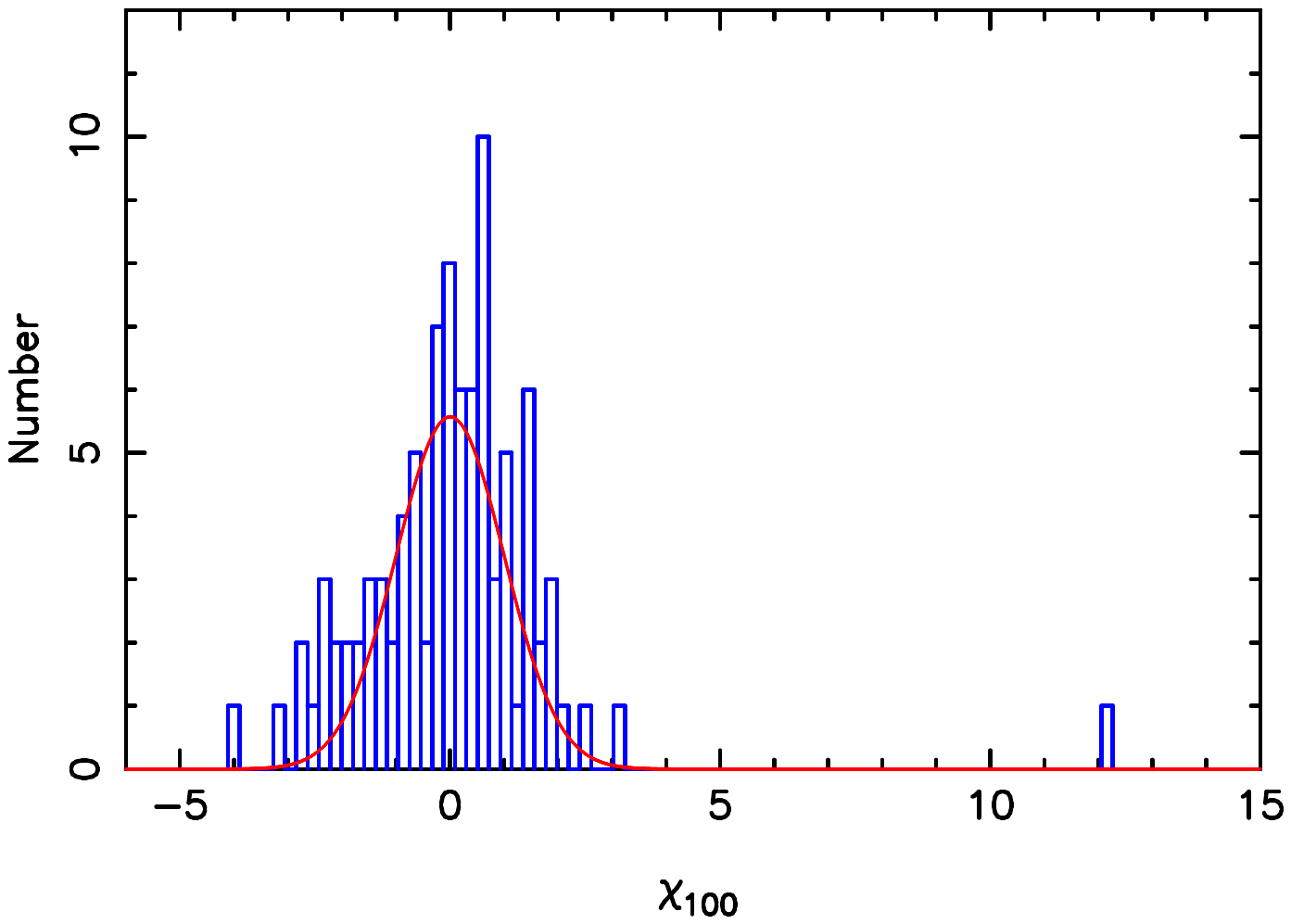} \includegraphics[angle=0] {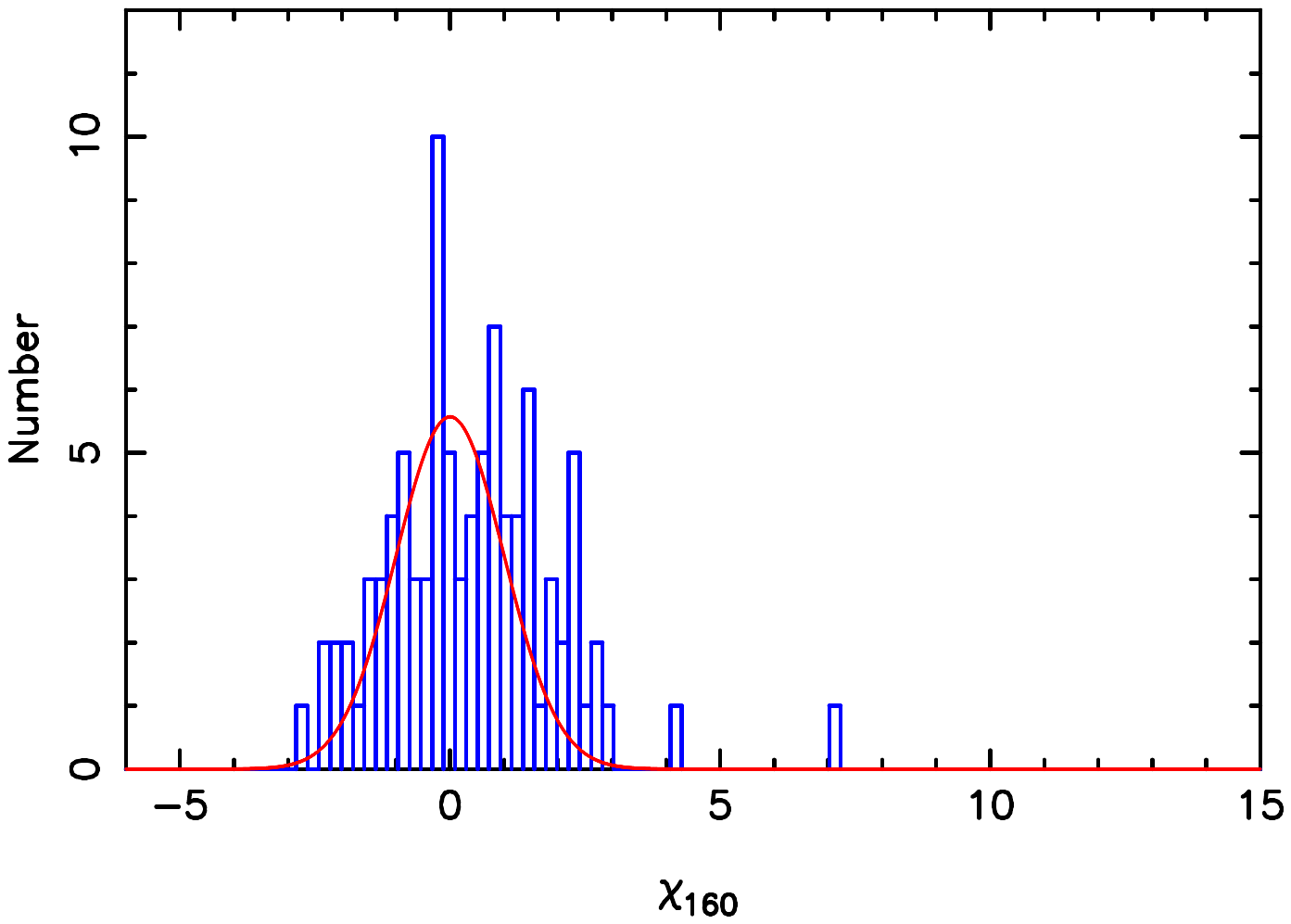}}
\caption{Distributions of the excess significances $\chi_{_\lambda}$ for the 94 DEBRIS
M-type dwarfs observed with {\it Herschel}/PACS at $\lambda$=100 and 160~$\mu$m.  The Gaussians shown in red are computed  
with $\sigma=1$  and normalized 
with the total number of stars excluding the two detections to serve as a  guide to characterize the observed noise distribution 
which is found to be well centered on zero. The excesses of the two disks GJ581 ($\chi_{100}=12.1$ and $\chi_{160}=4.2$) 
and Fom~C ($\chi_{100}=3.2$ and $\chi_{160}=7.2$) at 100 and 160~$\mu$m
are apparent in the plots.
}
\label{fig:chi}
\end{figure*}

\section{Disk incidence in the DEBRIS M-dwarf subsample observed with \textit{Herschel}/PACS} \label{incid}

We briefly summarize the procedures used to extract the relevant photometric data from the maps,  
referring to \citet{Kenn12a, Kenn12d} for a more detailed description.   
For the whole DEBRIS survey, the flux densities of the targets were measured using a combination
of point spread function (PSF) fitting and circular aperture photometry. 
{\it Herschel} calibration observations of bright stars were used as the PSF. They were rotated
to the angle appropriate for a given observation because the maps were created in sky coordinates.
The PSF fitting was used by default, with circular aperture photometry used where the PSF fitting residuals
revealed resolved sources. The preference for PSF fitting is largely
to allow for a mitigation of the effects of confusion from nearby point sources.
As the positions of the target stars were well known, the
PSF fitting routine was initialized at the expected star location
(or locations in the case of multiple systems). We then used the
MPFIT least-squares minimization {\it IDL} routine to find the best fitting
point source model for each observation.
For the aperture photometry, the uncertainty of the flux density  was estimated by applying the
same aperture at hundreds of randomly chosen positions in the map (see \citealt{Kenn12a}).
In the whole DEBRIS survey, the PSF fitted flux densities were  systematically
lower by $\sim$ 20 percent when compared to aperture photometry due to flux lost in the wings of the PSF
by filtering the images. Precisely, the typical aperture/PSF-fit flux density ratio based on  a
large number of targets was derived to be 1.19 at 100 $\mu$m, and 1.21
at 160 $\mu$m in \citet{Kenn12d}.

The flux densities thus measured ($F_{\lambda}$), photospheric levels ($P_{\lambda}$), 
and uncertainties ($e_{F_{\lambda}}, e_{P_{\lambda}}$) 
at wavelengths of $\lambda$ = 100 and 160~$\mu$m, for the 94 DEBRIS M dwarfs,
are listed in Table~\ref{tab:data}. This table also  gives the distances of the stars and  their
 effective temperatures from the stellar atmospheric  model
PHOENIX {\it Gaia} best-fit to stellar data from the literature \citep{Kenn12c,Kenn12d},
as in \citet{Thur14} and \citet{Sibt18} for the other DEBRIS subsamples. 
We achieved an approximately  uniform sensitivity of the survey at the levels of 2.0mJy at 100$\mu$m                                      
and 4.4 mJy at 160$\mu$m.
 
In total, significant flux densities at the star positions were found for nineteen M dwarfs in the  DEBRIS subsample 
($>3\sigma$). Possible excesses above their photospheric level were searched by computing the
flux excess significance of  $\chi_{_\lambda}$  at a wavelength of $\lambda$:

\begin{eqnarray}
\chi_{_\lambda}= {{F_{\lambda}-P_{\lambda}} \over {\sqrt{e_{F_{\lambda}}^2+e_{P_{\lambda}}^2}}}
\label{eq:chi}
\,
.\end{eqnarray}

\noindent The photometric data were taken from Table \ref{tab:data}, where
the $1\sigma$ PACS photometric uncertainties are $0.9~mJy < e_{F_{100\mu m}} < 3.2~mJy$ and
$2.6~mJy < e_{F_{160\mu m}} < 6.7~mJy$ depending on the target. 
The distributions of $\chi_{100}$ and $\chi_{160}$ 
for the M-dwarf subsample are shown in Fig.~\ref{fig:chi} and are approximately Gaussian 
and, satisfactorily, well centered on zero, implying that photospheric flux density estimates 
based on the PHOENIX stellar atmospheric model are unbiased. 
In DEBRIS, a disk is considered detected  when $\chi_{_\lambda}$ is larger than  3 at one or more wavelengths. 
According to this criterion, two debris disks were detected around the M3V star GJ~581 
and the M4V star Fomalhaut~C (Fom~C, LP~876-10) in the M-dwarf subsample.

 A slight excess on the positive side of both distributions in  Fig.~\ref{fig:chi} is apparent when compared 
with the Gaussian computed for pure noise. This slight excess is an indication of a possible population 
of disks just below the detection criterion ($\chi_{\lambda} > 3$ at one or both $\lambda$). In fact, quadratically 
combining $\chi_{100}$ and $\chi_{160}$ when both are positive in Table \ref{tab:data} yields three disks with 
$3 < \sqrt{(\chi_{100}^2+\chi_{160}^2)} \le 3.5$. We did not retain these as detections because this less conservative 
detection criterion was not adopted in the analysis of the other DEBRIS subsamples that we use for 
comparison in this paper.

The disk around GJ581 is resolved in the PACS images and modeling  
indicates a slightly extended disk with a resolved inner radius of $25\pm12$~au and a dust temperature 
significantly higher than blackbody, by a factor $f_T=3.5^{1.0}_{-0.5}$, at this radius  \citep{Lest12}. 
As introduced by \citet{Boot13}, this phenomenon can be cast into the ratio, $\Gamma=R_d/R_{BB}$, between the resolved disk radius, $R_d$ ,
and the fictitious blackbody radius, $R_{BB}$, resulting in a measured dust temperature of  $T_d=278.3 \times L_*^{0.25}/\sqrt{R_d/\Gamma}$. 
\citet{Pawe14} carried out a detailed study of 34 debris disks with two different modelings of the SED 
based on the  modified blackbody spectrum for one and on a grain size distribution for another. 
They found, for the former , $R_d$=38~au, along with $T_d=35\pm3$~K 
and $\Gamma=5.4\pm0.78$ (hence $R_{BB}=7$~au) for the disk around around GJ581, using DEBRIS {\it Herschel} photometry. 
As argued in \citet{Pawe14}, the latter model, with a grain size distribution, is likely more realistic
and provides a lower $\Gamma$. However, we report here the result of the modified blackbody model for consistency with the
published analysis of the other disk around an M-type DEBRIS star; namely, Fom~C (quoted in the next paragraph and used in our study as well).
Additionally, the fractional dust luminosity  derived from the modeled SED is $L_{dust}/L_*=8.9 \times 10^{-5}$ \citep{Lest12}.

The disk around Fom~C was resolved with ALMA after its discovery with {\it Herschel}/PACS 
\citep{Kenn14}. The ALMA image at 0.87~mm and its azimuthally averaged brightness profile
show a narrow belt with a radius of 26~au. The SED is modeled as a modified blackbody spectrum 
with photometry at 70, 100, 160, and 870~$\mu$m, and with  upper limits of 250, 350, and 500~$\mu$m. These results indicate a dust temperature of $20\pm4$~K, which is hotter than the blackbody temperature
at the resolved radius by the factor $1.4\pm0.2$ and so $\Gamma=2.0 \pm0.56,$ as analyzed in \citep{Cron21}. 
Hence, its blackbody radius, $R_{BB}$, is 13~au.  
The fractional dust luminosity derived from their modeled SED is $L_{dust}/L_*=1.5\pm0.2 \times 10^{-4}$.

\begin{figure}[h!]
\resizebox{10.0cm}{!}{\includegraphics[angle=0] {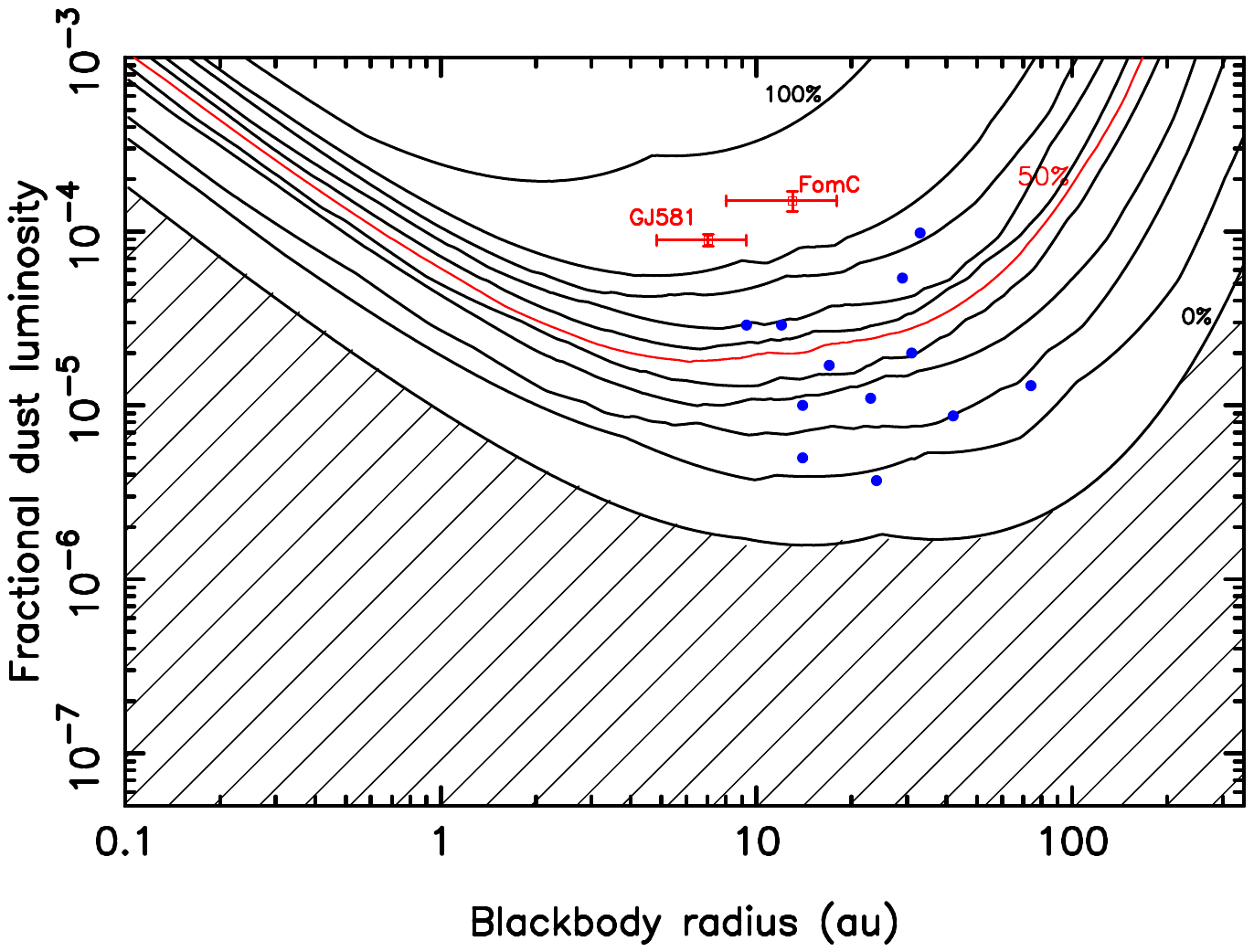}}
\caption{Completeness of the DEBRIS M-dwarf survey at $\lambda=100$ and 160~$\mu$m  
in the parameter space dust fractional luminosity versus black body radius. 
The contour lines show levels of completeness from zero to 100\%, in steps of
10\%. No detection is expected in the cross-hatched region and all disks 
are expected to be detected above the 100\% contour. 
The two disks discovered in the DEBRIS M-dwarf subsample are marked in red.
The detected disks of the DEBRIS K-star subsample in \cite{Sibt18} are marked 
as blue dots for comparison.
}
\label{fig:comp}
\end{figure}

\begin{figure}[h!]
\resizebox{8.cm}{!}{\includegraphics[angle=0] {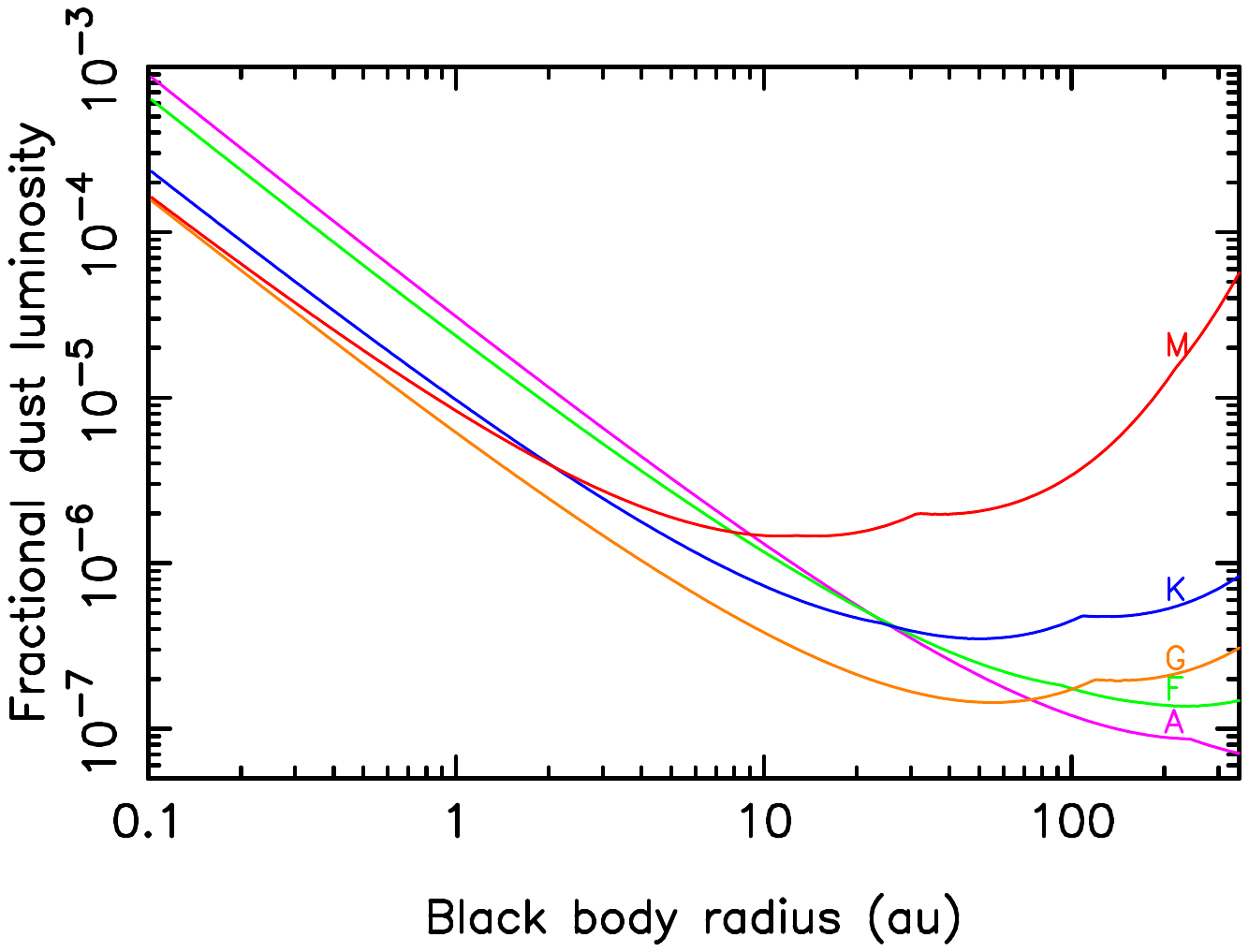}}
\caption{Detection threshold (0\% level of completeness) for the five DEBRIS subsamples A, F, G, K, and M, computed 
in using  their $100$ and 160~$\mu$m sensitivy threshold.  
}
\label{fig:zerocomp}
\end{figure}

Furthermore, we tried to improve the sensitivity of the survey to slightly resolved disks by using 
aperture photometry with an elliptical aperture, which is a better
match to a disk randomly oriented in space that is projected as an ellipse on the sky. 
For this purpose, we optimized the recovery of any extended emission centered on the star position
by computing the azimuthally averaged brightness profile from each image as a function of the semi-major axis, by incrementally varying the  ellipticity and position angle of the elliptical aperture. 
For each increment, we searched for any statistically significant deviation of the computed profile from the PSF 
that would be the signature of a resolved disk. We validated this search method 
by successfully recovering all previously
identified disks of other stellar types in DEBRIS, especially the nine resolved DEBRIS A-stars with 
their proper inclinations and orientations as modeled by \citet{Boot13}.
However, in the whole survey (AFGKM-stars), no new disk was discovered with this method to add to 
those already identified with the excess significance $\chi_{_\lambda}$ approach.  

Finally, two disks discovered among the 94 targets of the DEBRIS M-dwarf subsample
implies a detection rate of  $2.1^{+2.7}_{-0.7}$~\%, derived so that  the integrated 
probability of the rate between the lower and upper uncertainties is 68\%
(see the appendix of \citealt{Burg03} for a detailed description of
the computation of a low rate and its uncertainty based on the binomial distribution).
This can be compared  to the other rates within the whole DEBRIS program: 
$24 \pm 5$~\% for  A-stars \citep{Thur14} and 
$17^{+2.6}_{-2.3}$~\% for FGK-stars as estimated with the same method by \citet{Sibt18}  who
also noticed a trend across the three stellar types F, G, and K (see Table~\ref{tab:mod}).

\section{\textit{Herschel}/SPIRE observations}\label{SPIRE}

With the SPIRE observations, we surveyed  25  M dwarfs of the DEBRIS subsample to search for colder dust
in belts larger than usually expected. This resulted in no detection and a $3\sigma$ upper limits  of
$\sim$24 mJy for their flux densities at 250, 350, and 500~$\mu$m (see Table~\ref{tab:data}).
We derived a probability of 58\% for no detection in our SPIRE survey of 25 dwarfs,
using the binominal distribution with an average detection rate of $2.1\%$.
With these SPIRE observations, an upper limit of $3\times 10^{-4}$ can be set  on the fractional dust luminosity
of a belt as large as  $\sim$100~au (10~K for the mean stellar luminosity of 0.0084~L$_{\odot}$).

\begin{table*}[!h]
\caption{DEBRIS subsamples.}
\label{tab:mod}
\begin{tabular}{c c c c r r c r l}
\hline\hline
            \noalign{\smallskip}  
   SpT    & Targets  & $D_{MAX}$ & $<D_*>$  &  $<L_*>$        &  $<T_*>$ & Detections & Detection                 & References \\
          &          &  (pc)~~   & (pc)~    &  $(L_{\odot})$~ &  (K)~    &            & ~rate~~~~                 &   \\
            \noalign{\smallskip}
            \hline
            \noalign{\smallskip}
    A     &   86     &  45.5     &  32.0    &  27.2             &  8490    & 21        &  $24\pm 5$~\%              & {\citet{Thur14}} \\
          &          &           &         &                 &                      &                            &                             \\
    F     &   92     &  23.6     & 17.6    &  3.7            &  6415    & 22        &  $23.9^{+5.3}_{-4.7}$~\%   & {\citet{Sibt18}} \\ 
            &        &           &     &             &                &                      &                       \\
    G     &   91     &  21.3     & 16.2    &  1.11           &  5760    & 13        &  $14.3^{+4.7}_{-3.8}$~\%   & {\citet{Sibt18}} \\
            &        &           &     &             &                &                      &                       \\
    K     &   92     &  15.6     & 11.8    &  0.20           &  4610    & 12        &  $13.0^{+4.5}_{-3.6}$~\%   & {\citet{Sibt18}} \\
          &          &           &     &             &                &                      &                       \\
    M     &   94     &  8.6      &  6.2    &  0.012         &  3290    &  2        &   $2.1^{+2.7}_{-0.7}$~\%   & {This work}  \\    
            \noalign{\smallskip}
            \hline 
           \end{tabular} 
\end{table*}

\section{Completeness in the DEBRIS M-dwarf survey} \label{complete}

The ability to detect excess emission from a disk in the survey varies from target to target.
It depends on the properties of the disks (fractional dust luminosity, radius and dust composition) and 
of the stars (luminosity and  distance), as well as on the exact depth of the survey. 
Following \citet{Sibt18} for the other DEBRIS 
 subsamples, we evaluated this ability for the DEBRIS M-dwarf subsample in 
the 2D parameter space, $f_d$ versus $R_{BB}$, by using 
the same canonical model of a narrow, circular belt filled
with dust of fractional luminosity, $f_d$, and emitting as a blackbody at  a radius, $R_{BB}$, under the irradiation 
of the star at the center of the system. The flux density of this dust thermal emission is:

\begin{eqnarray}
  S_{\nu} = 2.95 \times 10^{19} \times B_{\nu}(\nu, T(R_{BB})) \times f_d \times R_{BB}^2 /d^2  ~~~~(mJy) 
\label{eq:Snu}
\,
,\end{eqnarray}

\noindent where the Planck function, $B_{\nu}$, is in W/m$^2$/Hz/sr, the distance to Earth, $d$, is in pc, and the radius,
$R_{BB}$, is in au and related to the dust temperature $T(R_{BB})$ through the standard relation in \S\ref{incid}.
For each point ($f_d$,~$R_{BB}$) of the diagram, we calculated the fraction of the stars having  disks that can be 
detected above the 3$\sigma$ level in the sample. This  3$\sigma$ level is taken to be the square of the quadratic sum of the
PACS photometric uncertainty and photosphere uncertainty for each target. 
Combining all detection fractions calculated at all points ($f_d$, $R_{BB}$) provides a measurement of the completeness 
of the survey within the 2D parameter space of $f_d$ versus $R_{BB}$. 

Figure~\ref{fig:comp} shows the resulting completeness for the DEBRIS M-dwarf subsample. 
The region above the upper curve of the plot is 100 percent complete; that is, had all stars 
of the survey been surrounded with disks characterized by parameters 
within this region, they would all have been detected with the $3\sigma$ sensitivity of the  PACS observations. 
In the plot, the completeness contours decrease in 10 percent steps down to the cross-hatched region, 
where  disks  cannot be detected around any of the stars in the subsample. The ample range of blackbody 
radii chosen between 0.1~au and 200~au 
for  Fig.~\ref{fig:comp} corresponds to blackbody dust temperature between 470~K and 4.9~K in the subsample. 
The upper temperature is for the M0-type dwarfs of the subsample $(L_*=0.08~L_{\odot})$. The lower value of 4.9~K is the asymptotic
temperature reached when the interstellar radiation field (ISRF) starts to dominate the M-dwarf stellar radiation field 
at the belt radius (see Appendix A of \citealt{Lest09}). The range of $f_d$ adopted in Fig.~\ref{fig:comp} 
is from the brightest known for FGK-star disks ($\sim 10^{-3}$) down to the faintest known, the  Kuiper Belt ($\sim 10^{-7}$). 

The two M-dwarf disks detected in the DEBRIS subsample have been added to Fig.~\ref{fig:comp}
($R_{BB}$ and $f_d$ of GJ581 and Fom~C in Sect. \ref{incid}). 
They lie above the 90\% completeness contour, indicating
that they had very little chance of being missed in the survey. 
We note that it is pointless to correct our measured M-dwarf detection rate for completeness
in this sample with only two detections at 90\%.

The comparison of the detection thresholds (0\% level of completeness) among the five DEBRIS subsamples A, F, G, K, M
in Fig.~\ref{fig:zerocomp} indicates that the M-dwarf survey is increasingly shallower than the other surveys 
in the $f_d - R_{BB}$ parameter space. This is the result of multiple factors; while the photometric depth 
of the observation is about the same for all DEBRIS subsamples, the sensitivity to disk emission is critically diminished 
for the M dwarfs  because of their very low stellar luminosities; this is despite their being closer to the Earth (see Table~\ref{tab:mod}).

In Fig.~\ref{fig:comp}, we   add the 12 disks detected in 
the DEBRIS K-star subsample \citep{Sibt18} to show their positions relative to the completeness contours
of the M-dwarf survey in parameter space of  $f_d - R_{BB}$. This mapping allows us to mimic the observations of the K-stars 
with a degraded  photometric sensitivity resulting in a shallower survey. The number of K-star disks detectable in these conditions
can be predicted by summing the percent contour levels at the positions of the K-star disks in this  diagram. 
This reduces the number of detected  K-star disks to 4.9 that we rounded off by 5 detections to calculate 
the rate of $5.4^{+3.4}_{-1.5}\%$  at the 68\% confidence level for the subsample of 92 K-stars in using
the same method as for the M-dwarf rate $2.1^{+2.7}_{-0.7}$~\% above. 
Hence, the two rates are statistically consistent within 1.1 times the quadratically combined uncertainties.
Thus, the statistics indicates that disks around M dwarfs appear not to be markedly less common than around K-stars.

\begin{figure*}[h!]
\resizebox{9.5cm}{!}{\includegraphics[angle=0] {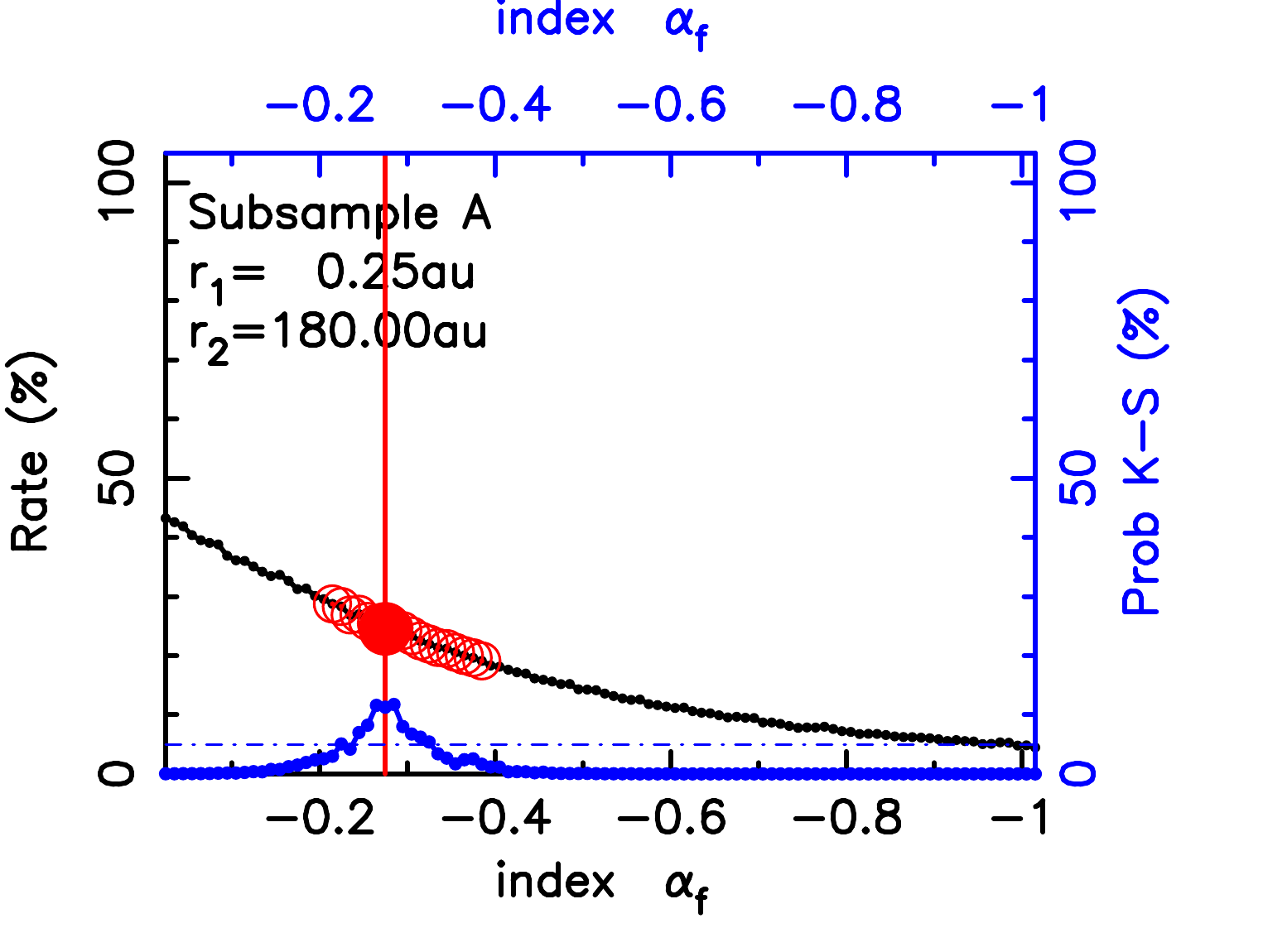} \includegraphics[angle=0] {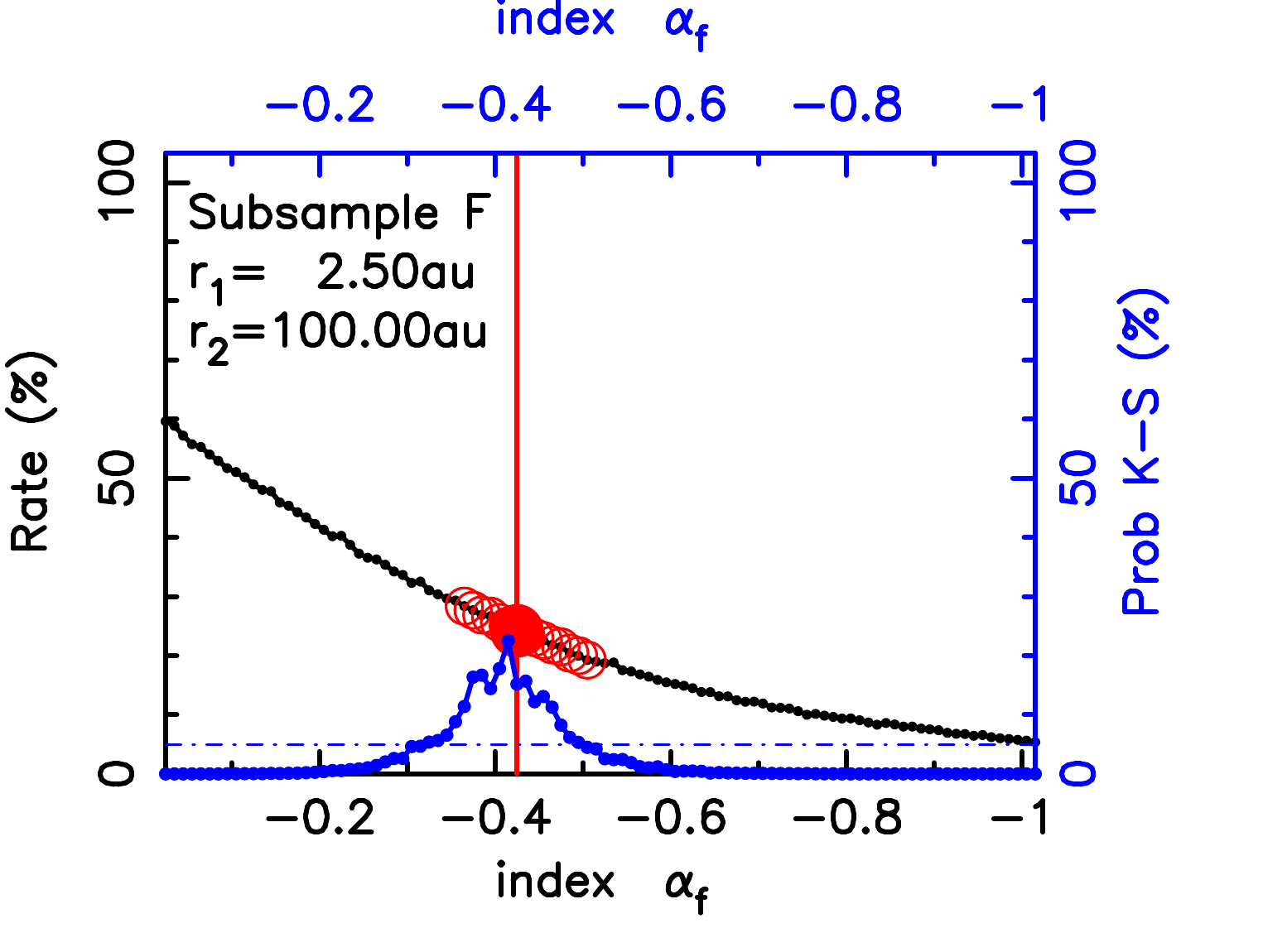} }
\resizebox{9.5cm}{!}{\includegraphics[angle=0] {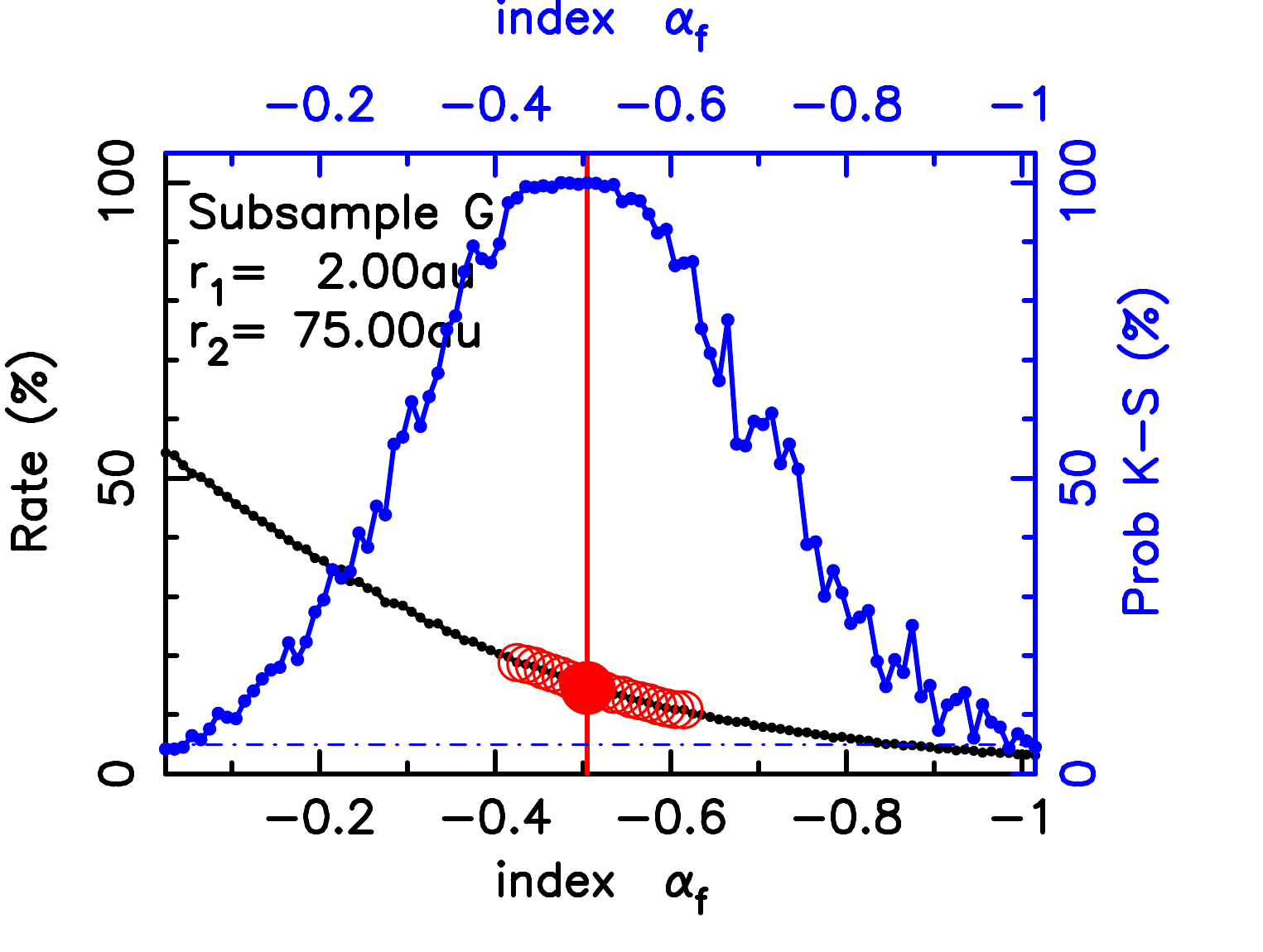} \includegraphics[angle=0] {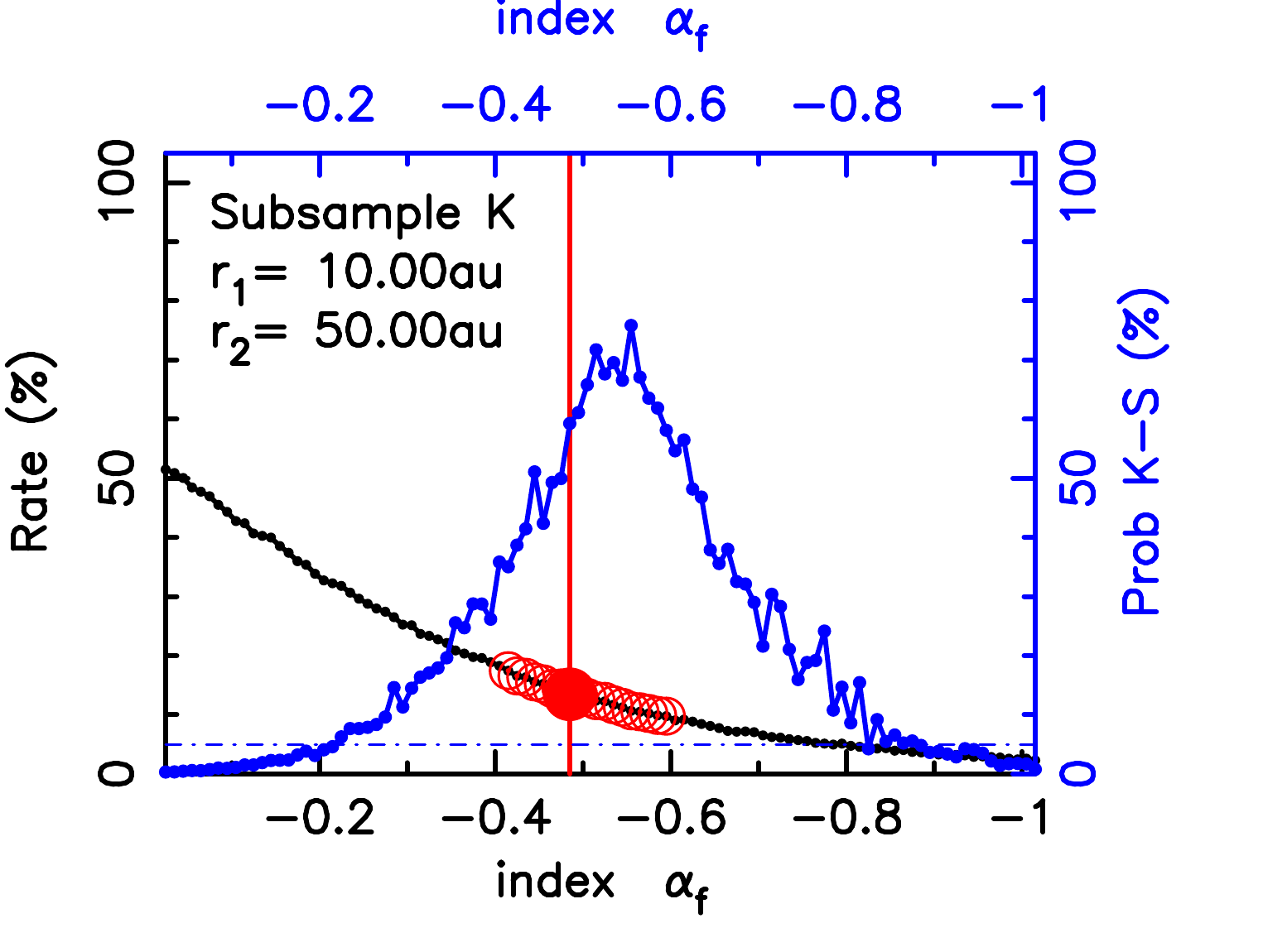} } 
\caption{Two-sample Kolmogorov-Smirnov test between the {\it Herschel}/PACS observations and the 
simulations of the DEBRIS subsamples A, F, G, and K. Each simulation assumes a power law of index of $\alpha_f$ 
for the distribution of the fractional dust luminosities of the disks. In blue, we show 
the Kolmogorov-Smirnov test significance level (Prob K-S in \%). The null hypothesis~(i.e., that the two samples are drawn 
from the same distribution) can be rejected if the significance level is below $5\%$ (horizontal dot-dash line). 
In black, connected dots are the simulated disk detection rates as a function of $\alpha_f$ for $r_1$ and $r_2$ of Table~\ref{tab:para}.
In red, the single filled large circle and vertical bar mark the observed detection rate and the unfilled large circles show its $1\sigma$ uncertainty range.
The K-S significance level peaks closely  at the same $\alpha_f$ as the observed detection rate for the four subsamples A,
F, G, and K, as expected with our fitting procedure and providing support for the power law model assumed.
The fit parameters are given in Table~\ref{tab:para}.
}
\label{fig:sim}
\end{figure*}

However, it is  striking that the only two M-dwarf disks discovered stand high in Fig.~\ref{fig:comp} with a detection probability of 
$>90\%$ in the subsample, whilst there is no other disk found in the diagram at a lower probability, unlike 
 the K-star disks in Fig.~\ref{fig:comp} and, more generally, unlike the FGK subsample in Fig.~4 of \citet{Sibt18}. 
It is an instructive exercise to limit the detections in the K-star subsample to the two brightest disks 
 and compute the probability of not detecting the ten other disks when mapped
into the M-dwarf parameter space of $f_d - R_{BB}$. Using  the positions of these ten K-star disks and completeness contours 
of Fig.~\ref{fig:comp}, this probability is  as low as  $0.9\%$ \footnote{The detailed calculation is : 
$(1-0.70)\times(1-0.68)\times(1-0.42)\times(1-0.40)\times(1-0.29)\times(1-0.27)\times(1-0.20)\times(1-0.18)\times(1-0.12)\times(1-0.08)$}. 
Thus, it is unlikely that the K-star survey would have detected only the two brightest disks. 
Hence, a caveat is that the two bright disks of the DEBRIS M-dwarf subsample found above the 90\% completeness contour are
possibly not statistically representative of the M-dwarf disk population and its true disk incidence could be lower; 
for instance, we have calculated with the binomial distribution that 
there is a probability of 50\% of not finding  a single disk in a sample of 94 stars 
 and a disk incidence of 0.75\%.  

\section{Constraints on the probability distributions of $f_d$ and $R_{BB}$ in the DEBRIS subsamples} \label{constrains}

\begin{figure*}[h!]
\resizebox{9.5cm}{!}{\includegraphics[angle=0,scale=1.0] {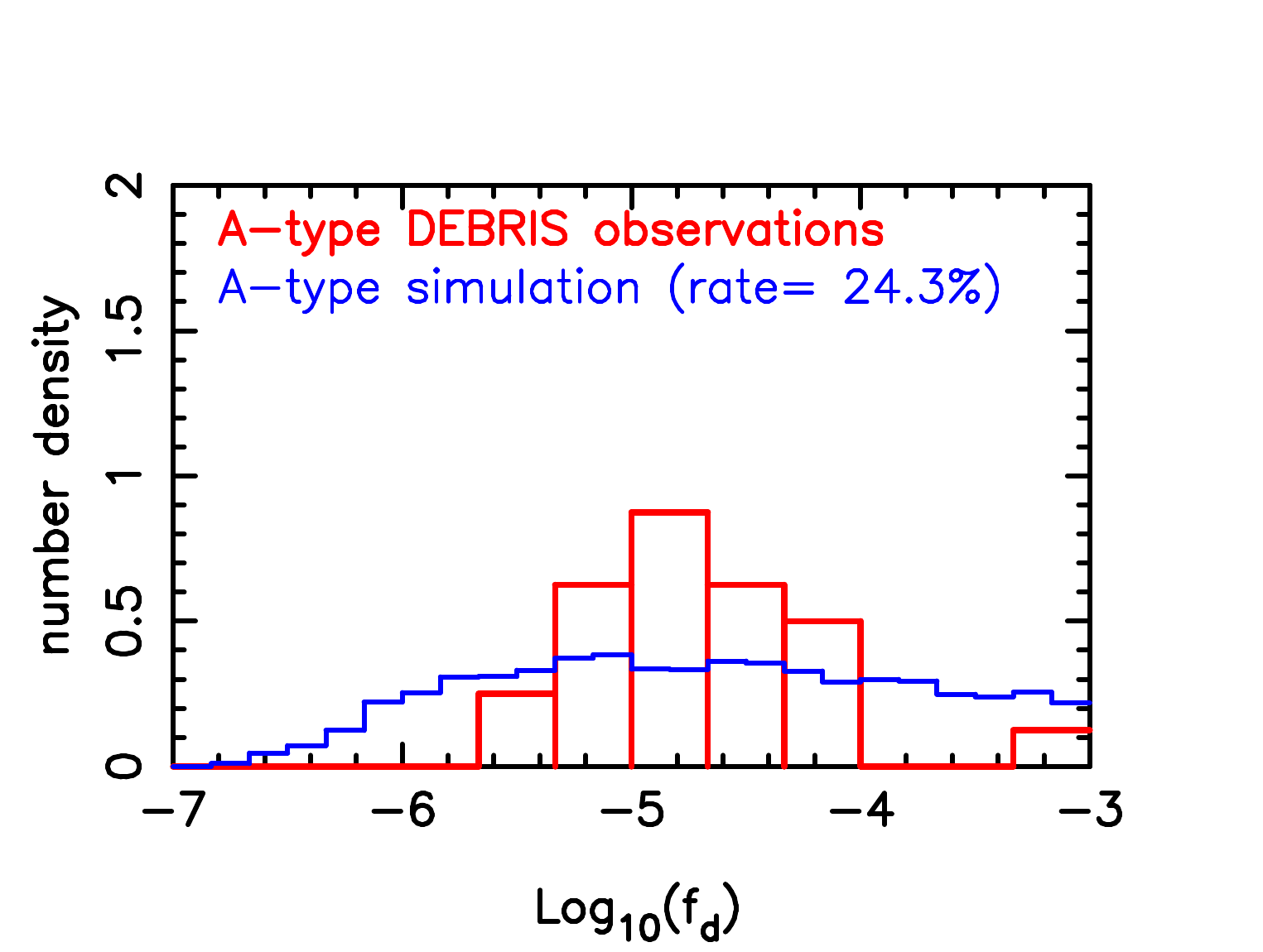} \includegraphics[angle=0,scale=1.0] {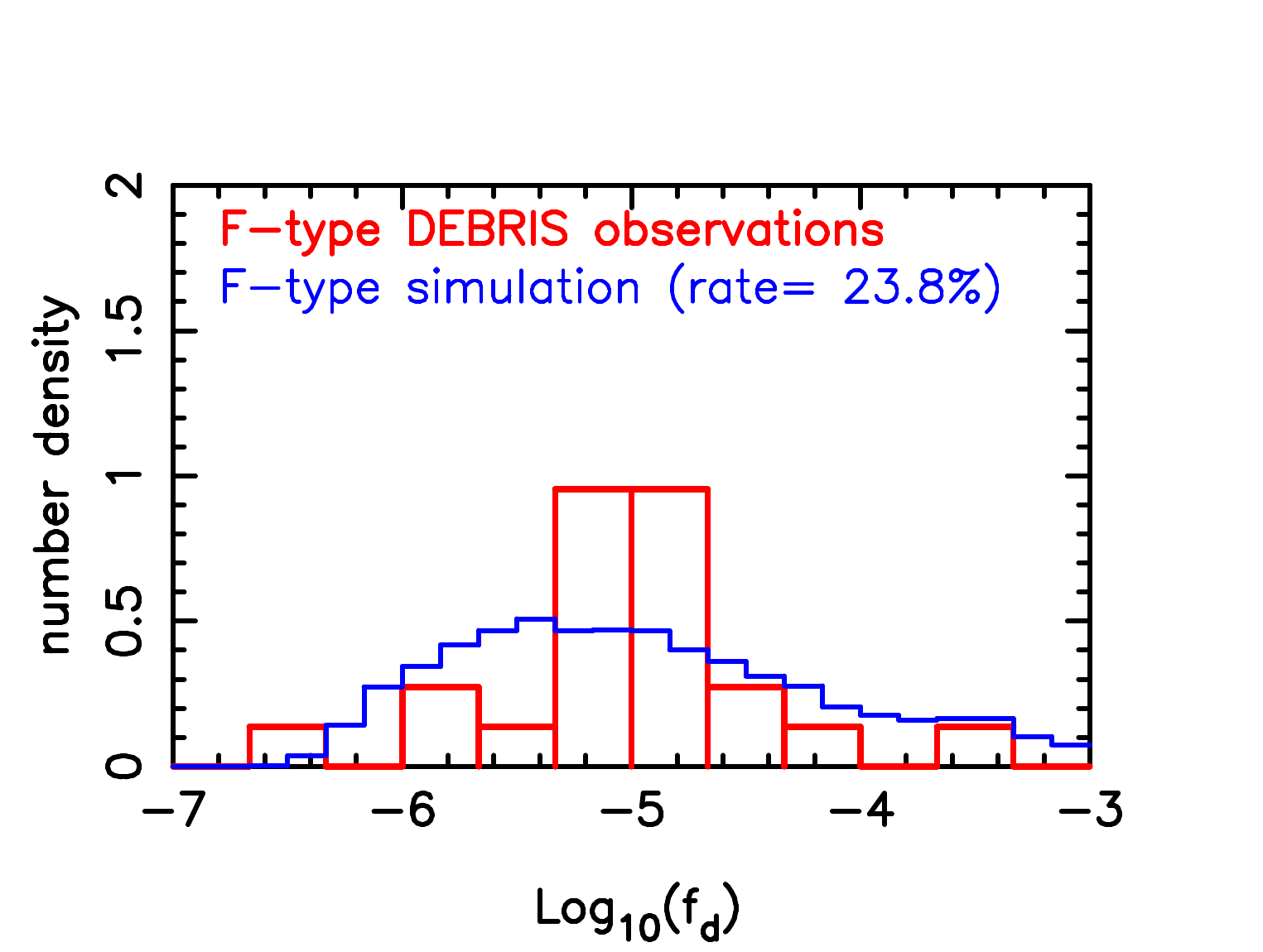}  } 
\resizebox{9.5cm}{!}{\includegraphics[angle=0] {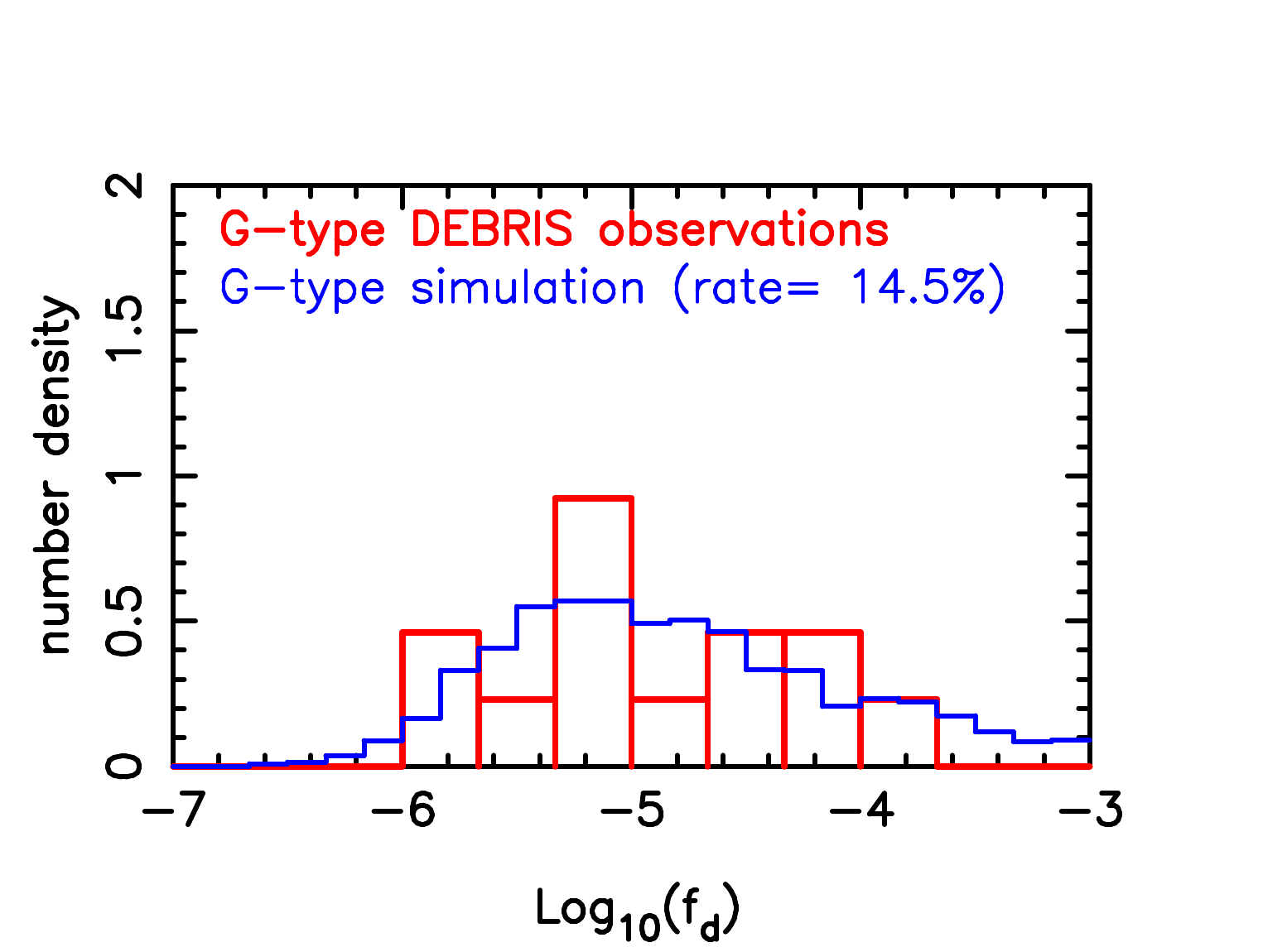} \includegraphics[angle=0] {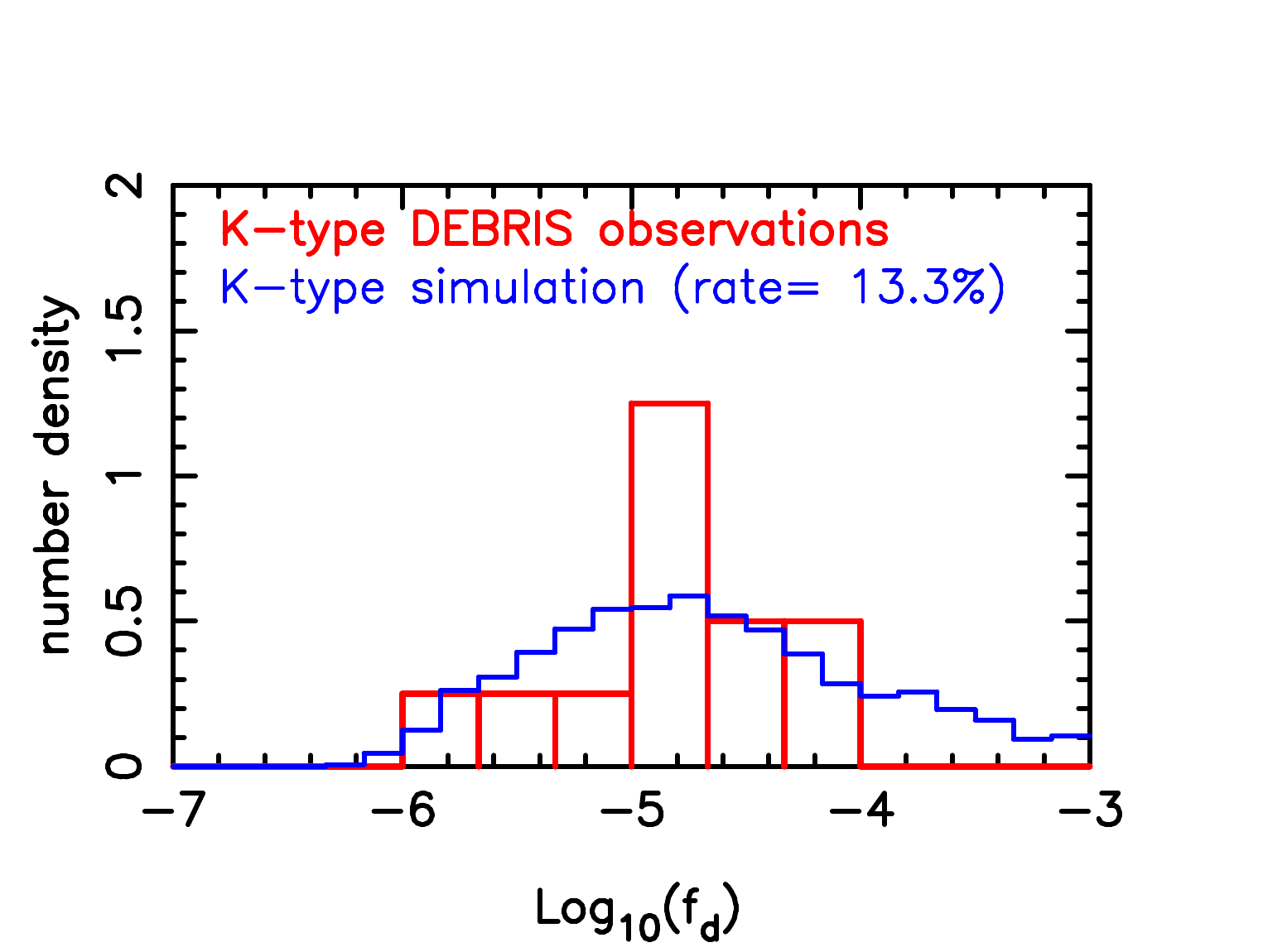}   }
\caption{
Observed $f_d$-distributions of the  detected disks for the DEBRIS subsamples A, F, G, K  
reported in \citet{Thur14} and \citet{Sibt18}, shown in red. The simulated $f_d$-distributions of the detected disks resulting from 
 our model and parameters in Table~\ref{tab:para} are shown in blue. 
The relatively smooth simulation histogram in blue is based on 250 independent realizations, while the observed histogram in red
on only one.
The detection rate in parenthesis is the mean from these 250 simulations (the observed detection rate is  reported in Table~\ref{tab:mod}).
The number densities have been normalized such that the total area under each histogram is unity.     
}
\label{fig:fd-dist}
\end{figure*}

We now explain how we derived constraints on the probability distributions of the two main parameters 
characterizing debris disks, fractional dust luminosities, $f_d$, and blackbody radii, $R_{BB}$, 
by using  the observations at 100~$\mu$m, the deepest of the survey. 
The build-up of a planetary system with its dusty belt(s) involves
complex underlying processes, along with the growth of planetesimals in a protoplanetary disk, planet migration, 
fragmentation of planetesimals through collisional evolution of the belt(s), and dust removal, which 
may lead to the specific system  properties (e.g.,
$f_d$ and $R_{BB}$) experiencing a scale invariance and thereby distributed as power laws \citep{Radi14}.

We have undertaken this hypothesis, assuming two power law distributed variates for  $f_d$ and $R_{BB}$ 
(${\rm d}N_f \propto [f_d]^{\alpha_f}~{\rm d}f_d$ and ${\rm d}N_r \propto [R_{BB}]^{\alpha_r}~{\rm d}R_{BB}$)
in the simple blackbody model of Eq.~\ref{eq:Snu}, to simulate the dust thermal emissions around all the DEBRIS stars.
In this model, the possible fractional dust luminosities are set between  $10^{-7}$ and $10^{-3}$,
while the  possible blackbody radii between the lower and upper bounds, $r_1$ and $r_2$, remain as free parameters.
In practice, the power law distributed variates $f_d$ and $R_{BB}$ between their lower and
upper bounds were generated from a uniformly distributed variate taken over [0,1],  using  the inversion method (\citet{Radi14} and
Wolfram~\footnote{https://mathworld.wolfram.com/RandomNumber.html}).
In our  simulations,  a disk is considered to be detected when its flux density  is
larger than three times the DEBRIS $100~\mu$m photometric uncertainty in Table~\ref{tab:data}.
We based our statistical study on a high number of  simulations.
 
First, we performed simulations of the subsamples A, F, G, K over a broad (but coarse) grid of values for 
all the free parameters $\alpha_f$, $\alpha_r$, $r_1$, and $r_2$ of the model. 
We found that the simulated disk detection rate and $f_d$-distribution of the detected disks are particularly 
sensitive to $\alpha_f$ (which  must be negative) and to the range $[r_1, r_2],$ which must be broad, but are not
markedly sensitive to $\alpha_r$ (which can be assumed zero).
Then, after setting  $\alpha_r$  to zero, we refined the values of the three parameters $\alpha_f$, $r_1$, and $r_2$ with 
the following fitting procedure. The observations that constrain these parameters are the observed disk detection rate and
distribution of the fractional dust luminosities of the detected disks in the subsample ($f_d$-distribution). 
We varied  $\alpha_f$  by small increments over $[0 , -3]$,  as well as $r_1$ over $[0.1, 30~\rm{au}]$ and $r_2$ over $[30, 300~\rm{au}]$
to simulate each subsample and find the values of the triplet ($\alpha_f$, $r_1$, $r_2$) so that
1) the simulated and observed disk detection rates are made to match and 2) 
the observed and simulated  $f_d-$distributions of the detected disks are drawn from the same distribution. This last
requirement is tested as the null hypothesis 
in the two-sample Kolmogorov-Smirnov test \citep[e.g.][]{Pres92}. 
The sizes of the two samples are suitable for the K-S test; the observation samples (A, F, G, and K) comprise 12 to 22 detected disks, while the simulation samples comprise hundreds of detected disks from the 250 simulations performed for each stellar type.
We adopted the standard criterion of 5\% for the K-S significance level (Prob K-S; Eq. 13.5.5 in \citealt{Pres92}); 
below this, the null hypothesis~(where the two samples are drawn from the same distribution) is rejected.
With this procedure, the solution  is when the
observed and simulated disk detection rates match for the same triplet ($\alpha_f$, $r_1$, $r_2$) as
the  K-S significance level (Prob~K-S) is maximized (at least as closely as possible in the case of subsample~K).
In Fig.~\ref{fig:sim}, we plot, as functions of $\alpha_f$, both
the simulated and observed disk detection rate and the K-S significance level for the optimum pair ($r_1$, $r_2$) found.
For each subsample, the resulting parameters of the solution are in Table~\ref{tab:para}. 
 
Finally, in Fig.~\ref{fig:fd-dist}, we show the observed and simulated $f_d-$distributions of the
detected disks for the solutions of Table~\ref{tab:para}. For each subsample, the
two distributions agree reasonably well in shape  (the K-S test in the
previous figures quantitifies this degree of
agreement). Differences are apparent but 
this is expected since the simulation histogram (blue) is based on 250 independent realizations while the observed histogram (red)
is based on only one.
 
To test our fitting procedure, 
we  generated many sets of mocked data for the stars of each DEBRIS subsample by adopting their luminosities, distances, 
and photometric uncertainties. We also set a~priori values for $r_1$, $r_2$, and $\alpha_f$ in the  model, including $\alpha_r=0$. 
All sets of mocked data were then analyzed with our procedure and we found that these a~priori values
 could be recovered within $\pm10\%$ and  $\pm50\%$ for  $\alpha_f$ and  $r_1$, respectively, 
and $^{+100\%}_{-50\%}$ for $r_2$. These relative uncertainties have been used to infer  
uncertainties of parameters in Table~\ref{tab:para}.  
In the course of these tests we also found that 
$r_1$ (as well as $r_2$) cannot be set to the same value for all four subsamples and so, these parameters must be intrinsic to each subsample with
our model and the DEBRIS data.  

We found that the K-S probability functions in Fig.~\ref{fig:sim} peak always well above the 5\% threshold but it can differ significantly between 
mocked datasets generated with the same values for $r_1$, $r_2$, and  $\alpha_f$, but with different variates of $R_{BB}$  and $f_d$. 
We also found that the K-S probability functions are 
significantly narrower when the photometric sensitivity is arbitrarily improved in the mocked data; namely, more disks are thus detected in the subsamples.
We found also that the K-S probability functions can be more or less
broad depending on the realizations and this must be a small number statistics effect.

From the resulting range of blackbody radii [$r_1$, $r_2$] of Table~\ref{tab:para}, 
we can  derive a mean resolved disk radius for each DEBRIS subsample in using the factor $\Gamma$ defined in Sect.\,\ref{incid}.
This factor can be evaluated with 
the empirical relation $\Gamma$-stellar luminosity found by \citet{Pawe14} in their extensive study of resolved disks observed from A to M-type 
stars, along with their complementary study of the impact of dust composition on this factor \citep{Pawe15}. These  studies predict
$\Gamma=$1.2, 1.6, 2.0, and 2.0 for the subsamples A, F, G, and K, respectively, using their mean stellar luminosities in Table~\ref{tab:mod}.
Thus, the  averaged,  resolved radii (${(r_1+r_2)/ 2} \times \Gamma$) calculated from Table~\ref{tab:para} are: 108, 82, 77, and 60~au. 
Although these four averaged values  estimated for subsamples A, F, G, and K  are uncertain (dominated by a 100\% uncertainty 
on $r_2$), they show a decreasing 
behavior with stellar luminosity that reminds of the  \citet{Matr18} empirical relation for mean resolved radii from the observations. 
However, we note that radius $r_1$, taken on its own in our modelling, shows the opposite, i.e. an increasing trend with stellar luminosity.

\begin{table}[!h]
\caption{Model parameters.}
\label{tab:para}
\begin{tabular}{c| c c c c }
\hline\hline
 \noalign{\smallskip}  
  Sample     &  $\alpha_f$               & $\alpha_r$  &  $r_1$             & $r_2$                \\
             &                           &             &   (au)             &  (au)                \\
\hline
             &                           &             &                    &                      \\
    A        &   $-0.28^{+0.10}_{-0.08}$ &   0         &    $0.25\pm0.1$    &  $180^{+180}_{-90}$    \\
             &                           &             &                    &                      \\
    F        &   $-0.43^{+0.10}_{-0.05}$ &   0         &    $2.5\pm1.2$     &  $100^{+100}_{-50}$    \\
             &                           &             &                    &                      \\
    G        &   $-0.51\pm0.10$          &   0         &    $2\pm1$         &  $75^{+75}_{-37}$      \\
             &                           &             &                    &                      \\
    K        &   $-0.49\pm0.10$          &   0         &    $10\pm5$        &  $50^{+50}_{-25}$      \\
             &                           &             &                    &                      \\
\noalign{\smallskip}
\hline 
\end{tabular} 
\end{table}

Finally, we turn to the DEBRIS M-dwarf subsample of  Table~\ref{tab:data}. The K-S test cannot be used because
 the observation sample comprises only two detected disks.  
So, we simply simulated the subsample by adopting the power-law model with the parameters of 
subsample K in Table~\ref{tab:para} ($\alpha_f=-0.49$, $\alpha_r=0.0$, 
$r_1=10$, and  $r_2=50$~au). Thus, we find that the simulated detection rate of the DEBRIS M-dwarf subsample
is  $3.1\pm1.9$\%, statistically consistent with the observed rate of $2.1^{+2.7}_{-0.7}$\%. 
We conclude from this that there is no evidence that the disk incidence drops
markedly between the M and K DEBRIS subsamples.

\section{Discussion}  \label{Dis}

This conclusion can be put in the context of several arguments 
that have led, instead, to postulate on lower dustiness in disks around M dwarfs than around FGK-stars;   
M-dwarf disks could be less gravitationally self-stirred 
because of less massive bodies embedded in them \citep{Keny01,Kriv18} 
or be less dynamically stirred by lower mass planets orbiting close to disk edge \citep{Wyat99,Pear14}; 
alternatively, they may have suffered a more severe depletion of their planetesimals during close stellar encounters  
when still embedded in the open clusters of their birth \citep{Lest11}. 
Also, M dwarfs are generally older  than earlier type stars, typically older than 1~Gyr, and their disks have
suffered collisional erosion over a longer duration.
To address this question, we  developed a population model for the disks of the DEBRIS A, F, G, K, and M subsamples, initially assuming
 power laws for the probability distributions of their fractional dust luminosities, $f_d$, and blackbody radii, $R_{BB}$, in the simple black body model for the dust emission.
With this model and the DEBRIS observations,  we have found that the range, $[r_1, r_2],$ for  uniformly distributed $R_{BB}$ 
is intrinsic to each subsample and the $f_d$-power law is moderately steep with 
an index $\alpha_f$ of about $-0.3$ for  subsample A and
about $-0.5$ for  subsamples F, G, and K. In our analysis, this latter value of the fractional dust luminosity 
index is also compatible with the DEBRIS~M subsample, but cannot be directly derived from the limited 
number of detected disks. 
At this stage, we have no evidence that  the DEBRIS M-dwarf disks are significantly less dusty. 

\begin{figure}[h!]
\resizebox{8.5cm}{!}{\includegraphics[angle=0] {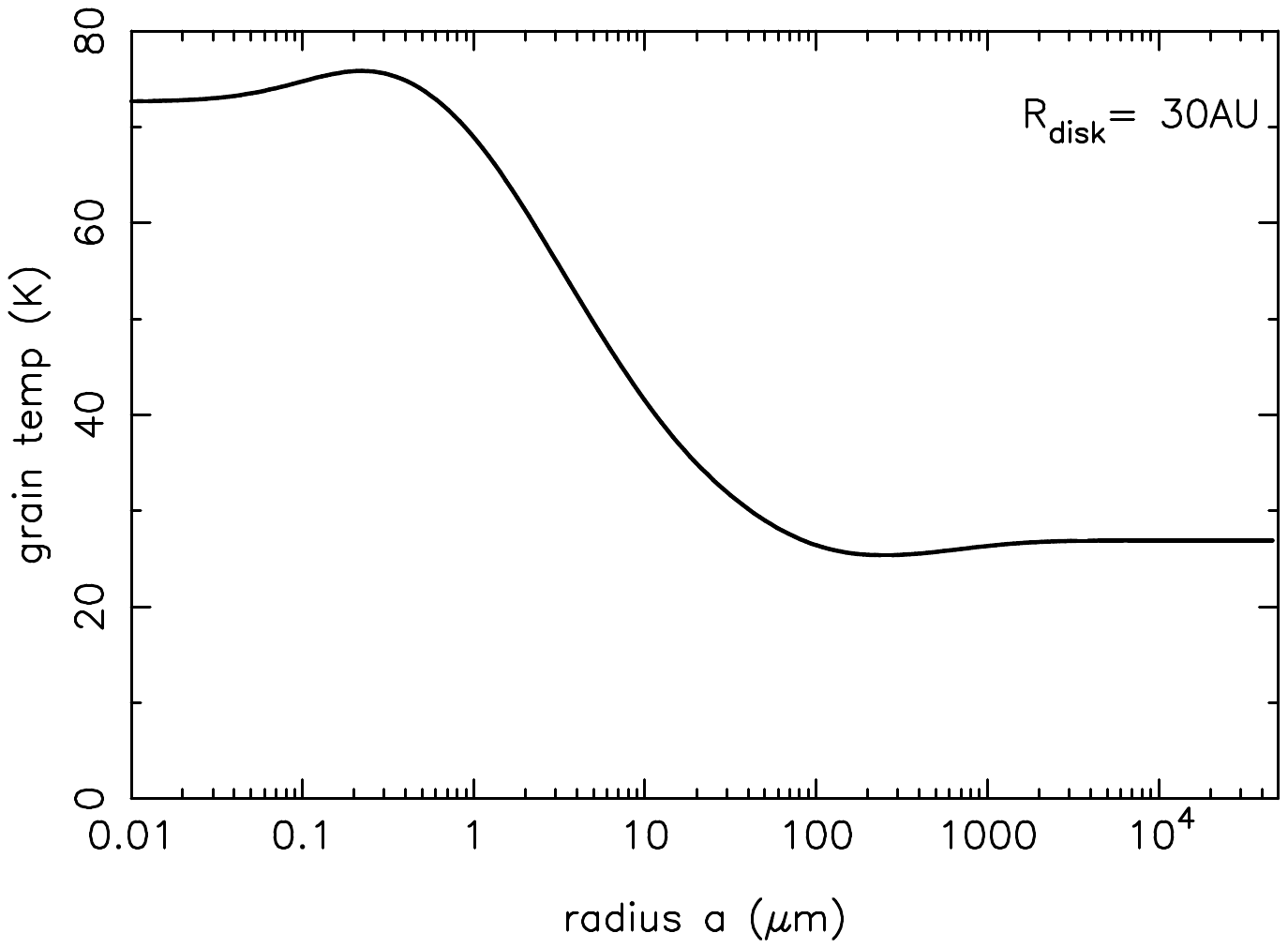}}
\caption{  
  Temperature of a spherical grain of {\it dirty ice} (water ice with embedded impurities of
  refractory materials of complex refractory index $m=1.33-0.09i$) located at 30~au from an M0-dwarf ($0.07~L_{\odot}$, 3850~K)
  and plotted as a function of its radius $a$. Large grains ($a > 100~\mu$m) are at the blackbody temperature of 26~K.} 
\label{fig:temp}
\end{figure}

\begin{figure}[h!]
\resizebox{8.5cm}{!}{\includegraphics[angle=0] {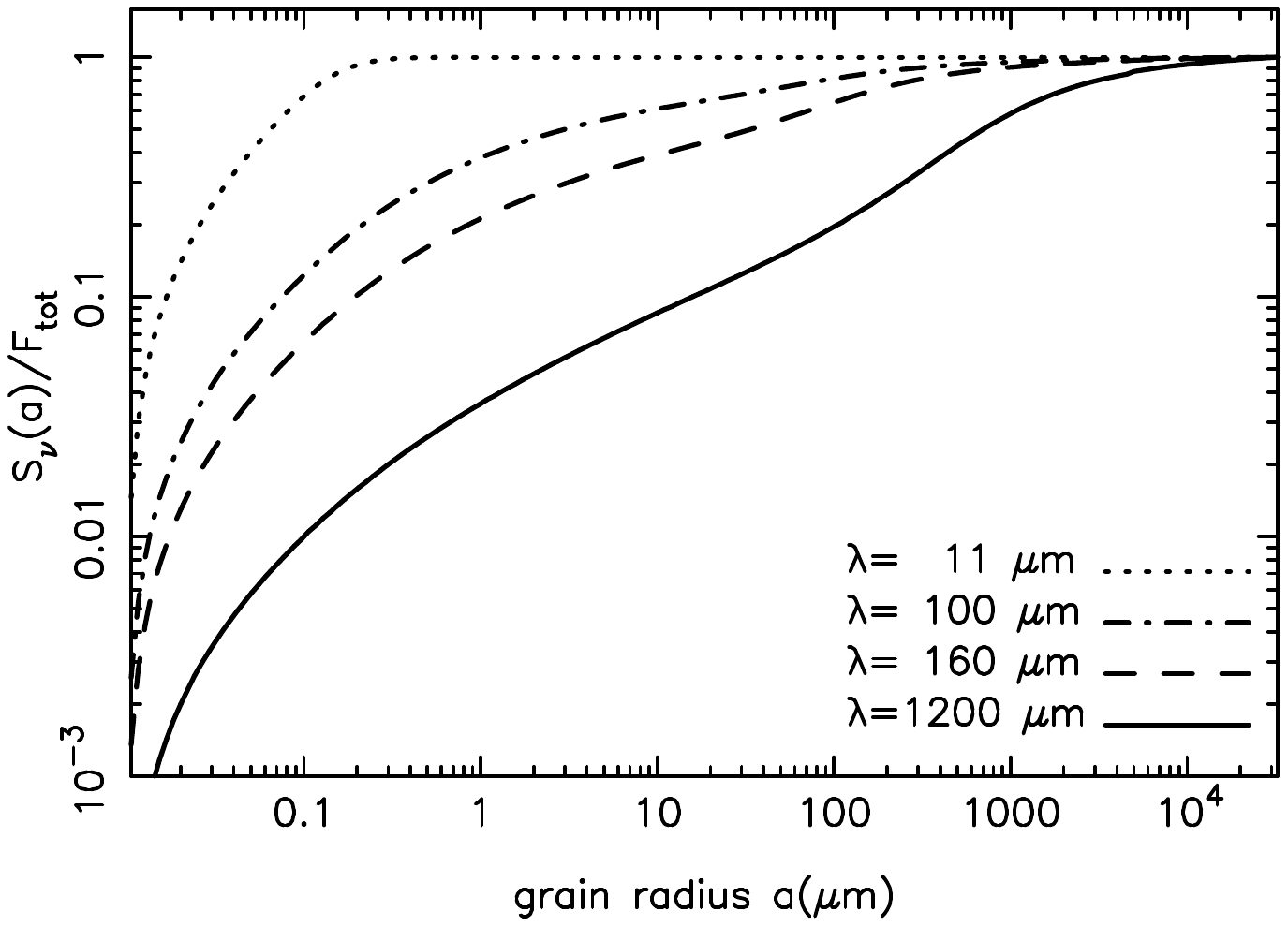}}
\caption{Cumulative contribution of different grain size intervals to
the flux density from emitting dust located at 30~au from an M0-dwarf (grain temperature in Fig.~\ref{fig:temp}). The {\it Herschel}/PACS observations
of the DEBRIS survey reported in this paper were  conducted at
$\lambda=$100 and 160~$\mu$m. As supplementary information, 
curves are provided for the early surveys at $\lambda=$11~$\mu$m by \citet{Plav05}  and at  $\lambda=$1200~$\mu$m by 
\citet{Lest09}. Flux density $S_{\nu}(a)$ integrated till grain size $a$ is normalized by 
the total flux density $F_{tot}$ integrated from minimum to maximum grain sizes of the plot.
}
\label{fig:dS-da}
\end{figure}

Another important question concerning the M dwarfs is the presence of an abundant population of
micron-sized grains  in their disks or not. 
On one hand, the insignificant radiation pressure around these low luminosity stars implies 
that even the smallest  grains cannot be expelled out of the system. On the other hand,
it has been argued that M dwarfs are magnetically active and have 
corpuscular winds with high mass loss rates that would expel dust grains smaller 
than a few $\mu$m out of the system \citep{Plav05}. However, it is now thought
that high magnetic activity is rather rare in M-dwarfs after \citet{Wood21} showed 
that the majority of them have winds similar in strength or even lower than the Sun.
The presence of an  abundant population of submicron-sized grains, necessarily with  temperatures
that are hotter than blackbody at the belt radius by a factor $2-3$ 
\citep{Back93},  would modify their  SED. 
As an illustration of this phenomenon, we compute in Fig.~\ref{fig:temp}
the temperature of a grain as a function of its size, $a$, for 
dust made of dirty ice  at a radial distance of 30~au from
an M0-type dwarf (see the method in Appendix~A of \citealt{Lest09}).
Then, using this temperature dependence and adopting 
the standard dust grain size distribution (${\rm d}N \propto a^{-3.5} {\rm d}a$ between a minimum size 
of 0.01~$\mu$m and a maximum size of 4 cm), 
we evaluate the cumulative contribution of different grain size intervals to the flux density  of this system in Fig.~\ref{fig:dS-da}. 
The impact of the minimum grain size on the flux density can be read directly from the plot. 
This shows that the flux densities at wavelengths 100 and 160~$\mu$m are  about halved 
if a minimum size of a few microns is taken and this would reduce disk detectability in the DEBRIS M-dwarf subsample. 
This plot shows also that nearly all the emission at $\lambda=11~\mu$m 
comes from grains smaller than 1~$\mu$m. Consequently, 
the dearth of debris disks in the small sample of nine young ($<500$~Myr) M dwarfs observed 
at this wavelength by \citet{Plav05} could be due to the dearth of these submicron-sized grains. They would have been blown out 
by high particle winds since in this case increased stellar magnetic activity is expected due to young age 
as suggested in their paper. Finally,
for the field M dwarfs of the DEBRIS survey, we note that the modeling of the two disks discovered does
show a dust temperature higher than  that of a blackbody (GJ581 in \citealt{Lest12} and Fom~C in \citealt{Cron21}),
making the presence of   an abundant population of hot micron-sized grains in these two disks plausible.

Several searches for debris disks around M dwarfs have been conducted in the far-infrared domain since 
the first debris disks were discovered with the satellite IRAS \citep{Auma84}.  
The {\it Spitzer} survey at 70~$\mu$m by \citet{Gaut07} searched a
sample of 41 nearby M dwarfs as part of a program to obtain, in an unbiased way, photometry of all stars lying
within 5 pc of the Sun. No excess was found with a sensitivity estimated to be about twice as worst as that of DEBRIS. 
If the DEBRIS disk detection rate of 2.1~\% is applied to this {\it Spitzer} sample, the probability of this outcome is 42\% 
and so, it can be statistically expected. The {\it Herschel}/PACS survey at $\lambda$=100 and 160~$\mu$m by \citet{Kenn18} searched 
a sample of 21 nearby M dwarfs 
hosting low-mass planets  and found a disk detection rate as high 
as 14\%~\footnote{We add that the uncertainties of this rate are $14.3^{+10.9}_{-4.6}~\%$ at a 68\% confidence level using the same methodology 
as in Sect. \ref{incid}.}. These authors carried out a statistical 
analysis that shows that this higher rate is more readily attributed to the superior depth of their survey (mean noise rms of 1.4 
and 3.6 mJy/b at 100 and 160~$\mu$m, respectively)    
than to a correlation between planets and debris disks. The {\it Herschel}/PACS survey at $\lambda$=100 and 160~$\mu$m 
by \citet{Tann20} searched a sample of 20 nearby ($<40$pc) and young ($<300$~Myr) late-type stars (19 M dwarfs
and one K7V type-star)  with a sensitivity similar to DEBRIS 
and found confidently one new debris disk around the K7V star TYC~9340-437-1,
resolved as a large belt with $R_d=96$~au ($\Gamma=4$) and with a high 
$L_{dust}/L_*$ of $1.2\times10^{-3}$ placing this late K-star  above the 100\% completeness contour in Fig. \ref{fig:comp}.      
It is important to note that this star is part of the young and nearby $\beta$~Pic moving group (20Myr).
The ALMA survey at $\lambda$=880~$\mu$m by \citet{Cron23} searched 36 M dwarfs all selected in the $\beta$~Pic Moving Group
and found a disk detection rate of $11.1^{+7.4}_{-3.3}$~\% significantly larger 
than in our unbiased  DEBRIS M-dwarf subsample. We note that their selection criterion seems to be determinant since 
 it  has led to a $\sim$75\% detection rate of debris disks around
F-type stars in the $\beta$~Pic moving group \citep{Pawe21}, while a rate of  $23.9^{+5.3}_{-4.7}$~\% 
is measured for the unbiased DEBRIS F-star subsample \citep{Sibt18} 
corroborating the fact (well established now) that youth of the system significantly favors detectability.  

The low disk detection rate we measured in the unbiased DEBRIS M-dwarf subsample must  
be attributed to the depth of our  {\it Herschel}/PACS observations. To  
further investigate the incidence of disks around M dwarfs,  
a  new, large, unbiased survey with ALMA with a significant improvement in depth 
and number of targets is needed but it currently requires a prohibitively long observing time \citep{Lupp20}.
Loss of sensitivity to large-scale emission with ALMA for the broadest disks  may be 
a limitation however. Thus, the first step is
to conduct a large unbiased survey capable of detecting the full variety of 
faint disks expected around M dwarfs with a new, large, single-dish telescope, 
such as the one proposed by the Atacama Large Aperture Submillimeter Telescope (AtLAST) project 
\citep{Holl19, Klaa24}.

\section{Conclusion}  \label{Con}

Within the {\it Herschel} Open Time Key program DEBRIS, we searched for debris disks around the 94 nearest M dwarfs
observed optimally in the far~infrared domain to detect cold dust emission 
with the {\it Herschel}/PACS observatory. This is the deepest, unbiased survey
of low-mass stars to date. Two debris disks (GJ581 and Fomalhaut~C) have been discovered and already reported 
in the literature.  The resulting low disk detection rate of $2.1^{+2.7}_{-0.7}$~\%  
is significantly less than in the other DEBRIS unbiased subsamples~: $17.1^{+2.6}_{-2.3}$~\% for the FGK subsample \citep{Sibt18} 
and 24$\pm$5\% for the A subsample  \citep{Thur14}. However, we show that
the DEBRIS M-dwarf survey is about ten times shallower than the DEBRIS FGK survey 
 in the  dust fractional dust luminosities versus blackbody radii parameter space. 
We argue  that, had the DEBRIS K-star subsample been observed at the same shallower depth in this physical 
parameter space, then its measured disk detection rate would have been statistically consistent with the one found 
for the M-dwarf subsample. Hence, in the DEBRIS survey, the incidence of debris disks does not appear 
to drop from the  K to M  subsamples when considering disks in the same
region of the physical parameter space and despite the decline of their observed detection rates. 
One caveat is that since the two M-dwarf disks discovered in the survey are standing alone above the 90\% completeness contour 
in the $R_{BB}-f_d$ plane, they may not be 
statistically representative and, thus, they would not be applicable to studies aimed at properly deriving the disk incidence of the whole population. 

\section{Data availability}

The full version of Table~\ref{tab:data} is available in the online arXiv version of this paper and also 
available in electronic form at the CDS via anonymous ftp to cdsarc.u-strasbg.fr (130.79.128.5) 
or via http://cdsweb.u-strasbg.fr/cgi-bin/qcat?J/A+A/. Star properties, photometric data from \textit{Herschel}/PACS and {\it Herschel}/SPIRE, 
and excesses above photospheric level  used  in this paper are provided.

\begin{acknowledgements}
We are in debt to the anonymous referee of our paper for providing insightful comments. 
The \textit{Herschel} spacecraft was designed, built, tested, and launched under a contract to ESA managed 
by the \textit{Herschel}/\textit{Planck} Project team 
by an industrial consortium under the overall responsibility of the prime contractor Thales Alenia Space (Cannes), 
and including Astrium (Friedrichshafen) responsible for the payload module and for system testing at spacecraft level, 
Thales Alenia Space (Turin) responsible for the service module, and Astrium (Toulouse) responsible for the telescope, 
with in excess of a hundred subcontractors. 
JFL  gratefully acknowledges  the financial support of Centre National d'Etudes Spatiales (CNES).
M.B. received funding from the European Union’s Horizon 2020 research and innovation program under 
grant agreement no. 951815 (AtLAST) and from the Deutsche Forschungsgemeinschaft (DFG) through grant Kr 2164/13–2. 
\end{acknowledgements}
  
\bibliographystyle{aa}

\bibliography{aa51673-24corr_JFLverified}

\onecolumn   
\begin{landscape}
\begin{longtable}{lllllrrrrrrrc}
\caption{{\it Herschel} photometry of the DEBRIS M-dwarfs.}  \\
\label{tab:data} \\
\hline\hline
DEBRIS    & Name         & HIP or HD   &  Spec.   &  $T_{eff}$ & dist & $F_{100}$~~~~  & $P_{100}$~~~~~  &   $\chi_{100}$   & $F_{160}$~~~~   &   $P_{160}$~~~~~~  &   $\chi_{t160}$  &  SPIRE  \\    
name      &              & number      &  type    &  (K)       & (pc) & (mJy)~~~       &  (mJy)~~~~      &                  & (mJy)~~~        &   (mJy)~~~~        &                  &  (mJy)  \\
\hline
\endfirsthead
\caption{continued.}\\ 
\hline
\endhead
\hline
\endfoot         
M002A     & GJ 406       &              &  M5.5   &  2656  &   2.39 &    4.8  $\pm$    1.9 &   3.77  $\pm$   0.06 &   0.54 &    5.0  $\pm$    4.1 &   1.63  $\pm$   0.03 &   0.83 & $<24$ \\
M003A     & GJ 411       & HD 95735     &  M2     &  3530  &   2.54 &   29.5  $\pm$    2.6 &  29.55  $\pm$   0.38 &  -0.02 &   10.0  $\pm$    6.7 &  11.38  $\pm$   0.15 &  -0.20 & $<24$ \\
M004AB    & GJ 65A       &              &  M5.5   &  2972  &   2.68 &    7.2  $\pm$    2.1 &   6.29  $\pm$   0.08 &   0.43 &    4.3  $\pm$    4.9 &   2.42  $\pm$   0.03 &   0.38 & $<24$ \\
M006A     & GJ 905       &              &  M5     &  3054  &   3.16 &    4.0  $\pm$    2.0 &   3.65  $\pm$   0.31 &   0.19 &    0.4  $\pm$    4.5 &   1.41  $\pm$   0.12 &  -0.23 & $<24$ \\
M007A     & GJ 887       & HD 217987    &  M0.5   &  3647  &   3.28 &   28.1  $\pm$    2.6 &  26.52  $\pm$   3.34 &   0.38 &   15.4  $\pm$    4.5 &  10.22  $\pm$   1.29 &   1.10 & $<24$ \\
M008A     & GJ 447       & HIP 57548    &  M4     &  3053  &   3.35 &    0.1  $\pm$    2.1 &   5.11  $\pm$   0.16 &  -2.39 &    9.6  $\pm$    5.7 &   1.98  $\pm$   0.06 &   1.33 & $<24$ \\
M009A     & GJ 866AB     &              &  M5.5   &  2692  &   3.45 &    5.3  $\pm$    1.8 &   5.13  $\pm$   0.06 &   0.12 &   -1.8  $\pm$    3.6 &   2.26  $\pm$   0.03 &  -1.10 & $<24$ \\
M010A     & GJ 725A      & HD 173739    &  M3     &  3420  &   3.52 &   12.1  $\pm$    1.2 &  11.46  $\pm$   0.20 &   0.51 &    2.4  $\pm$    2.6 &   4.42  $\pm$   0.08 &  -0.77 & $<24$ \\
M010B     & GJ 725B      & HD 173740    &  M3.5   &  3082  &   3.52 &   11.9  $\pm$    1.2 &  12.44  $\pm$   1.31 &  -0.30 &    6.9  $\pm$    2.6 &   4.87  $\pm$   0.51 &   0.77 &       \\
M011A     & GJ 15A       & HD 1326      &  M2     &  3532  &   3.58 &   14.8  $\pm$    2.3 &  16.22  $\pm$   0.18 &  -0.60 &   13.2  $\pm$    3.2 &   6.27  $\pm$   0.07 &   2.20 & $<24$ \\
M011B     & GJ 15B       &              &  M3.5   &  3107  &   3.58 &    3.8  $\pm$    1.9 &   3.62  $\pm$   0.07 &   0.09 &    2.7  $\pm$    3.3 &   1.48  $\pm$   0.03 &   0.36 &       \\
M012A     & GJ 1111      &              &  M6     &  2610  &   3.63 &    1.7  $\pm$    2.1 &   1.19  $\pm$   0.01 &   0.25 &    3.7  $\pm$    4.0 &   0.53  $\pm$   0.01 &   0.78 & $<24$ \\
M013A     & GJ 1061      &              &  M5.5   &  2836  &   3.68 &   -0.6  $\pm$    1.9 &   1.98  $\pm$   0.02 &  -1.35 &   -7.0  $\pm$    3.8 &   0.85  $\pm$   0.01 &  -2.06 & $<24$ \\
M014A     & GJ 54.1      & HIP 5643     &  M4.5   &  3024  &   3.72 &    6.2  $\pm$    2.0 &   2.32  $\pm$   0.03 &   1.94 &    4.8  $\pm$    3.8 &   0.94  $\pm$   0.01 &   1.01 & $<24$ \\
M015A     & GJ 273       & HIP 36208    &  M3.5   &  3247  &   3.79 &    7.7  $\pm$    2.4 &   8.51  $\pm$   0.08 &  -0.32 &    3.7  $\pm$    5.2 &   3.29  $\pm$   0.03 &   0.08 & $<24$ \\
M017A     & GJ 191       & HD 33793     &  M1.0   &  3673  &   3.91 &    4.6  $\pm$    1.9 &   6.09  $\pm$   0.07 &  -0.80 &    1.7  $\pm$    3.2 &   2.35  $\pm$   0.03 &  -0.21 & $<24$ \\
M021A     & GJ 1         & HD 225213    &  M1.5   &  3644  &   4.34 &    4.2  $\pm$    2.0 &   9.80  $\pm$   0.19 &  -2.75 &   12.6  $\pm$    6.2 &   3.78  $\pm$   0.07 &   1.43 & $<24$ \\
M022AB    & GJ 473A      &              &  M5     &  3026  &   4.39 &    3.9  $\pm$    2.1 &   3.30  $\pm$   0.10 &   0.28 &    0.2  $\pm$    3.5 &   1.31  $\pm$   0.04 &  -0.33 & $<24$ \\
M023A     & GJ 83.1      &              &  M4.5   &  3054  &   4.45 &   -2.3  $\pm$    1.7 &   1.85  $\pm$   0.05 &  -2.40 &   -7.6  $\pm$    3.4 &   0.74  $\pm$   0.02 &  -2.42 & $<24$ \\
M024A     & GJ 687       & HIP 86162    &  M3     &  3405  &   4.54 &    6.1  $\pm$    1.8 &  10.37  $\pm$   0.09 &  -2.41 &    0.2  $\pm$    3.5 &   3.99  $\pm$   0.03 &  -1.06 & $<24$ \\
M025A     & GJ 3622      &              &  M6.5   &  2886  &   4.54 &   -4.8  $\pm$    1.8 &   0.62  $\pm$   0.02 &  -3.08 &    0.3  $\pm$    3.1 &   0.26  $\pm$   0.01 &  -0.00 & $<24$ \\
M028A     & GJ 876       & HIP 113020   &  M4     &  3210  &   4.69 &    6.5  $\pm$    2.3 &   7.24  $\pm$   0.08 &  -0.31 &    6.5  $\pm$    4.0 &   2.80  $\pm$   0.03 &   0.92 & $<24$ \\
M029A     & GJ 1002      &              &  M5.5   &  3041  &   4.69 &    1.9  $\pm$    2.0 &   0.94  $\pm$   0.03 &   0.47 &   -3.4  $\pm$    3.5 &   0.38  $\pm$   0.01 &  -1.08 &       \\
M031A     & GJ 412A      & HIP 54211    &  M0.5   &  3640  &   4.86 &    6.3  $\pm$    2.1 &   7.74  $\pm$   0.14 &  -0.67 &    7.6  $\pm$    5.4 &   2.98  $\pm$   0.06 &   0.85 &       \\
M031B     & GJ 412B      &              &  M6     &  2801  &   4.86 &    0.1  $\pm$    1.9 &   0.75  $\pm$   0.01 &  -0.36 &    8.9  $\pm$    5.4 &   0.32  $\pm$   0.01 &   1.58 &       \\
M032A     & GJ 388       &              &  M3     &  3575  &   4.89 &    1.0  $\pm$    2.1 &   9.47  $\pm$   0.22 &  -4.02 &    4.6  $\pm$    3.7 &   3.65  $\pm$   0.08 &   0.27 &       \\
M033A     & GJ 832       & HD 204961    &  M1.5   &  3614  &   4.95 &   12.5  $\pm$    1.9 &   9.92  $\pm$   0.12 &   1.36 &    1.2  $\pm$    3.5 &   3.82  $\pm$   0.05 &  -0.75 &       \\
M036AB    & GJ 1116A     &              &  M5.5   &  2904  &   5.23 &    3.3  $\pm$    1.7 &   1.62  $\pm$   0.14 &   0.98 &    5.3  $\pm$    4.4 &   0.67  $\pm$   0.06 &   1.06 &       \\
M038A     & GJ 3323      &              &  M4     &  3064  &   5.32 &    2.7  $\pm$    1.8 &   1.75  $\pm$   0.14 &   0.55 &   -4.7  $\pm$    3.8 &   0.70  $\pm$   0.06 &  -1.42 &       \\
M039A     & GJ 445       & HIP 57544    &  M3.5   &  3520  &   5.34 &    3.2  $\pm$    3.1 &   2.98  $\pm$   0.17 &   0.06 &   -3.3  $\pm$    3.5 &   1.15  $\pm$   0.06 &  -1.27 &       \\
M040A     & GJ 526       & HD 119850    &  M1.5   &  3766  &   5.39 &    8.0  $\pm$    2.1 &   9.45  $\pm$   0.13 &  -0.70 &    3.5  $\pm$    5.1 &   3.65  $\pm$   0.05 &  -0.02 & $<24$ \\
M042A     & GJ 251       & HD 265866    &  M3     &  3322  &   5.61 &    5.2  $\pm$    2.2 &   5.58  $\pm$   0.09 &  -0.19 &    1.7  $\pm$    3.7 &   2.16  $\pm$   0.04 &  -0.12 &       \\
M044A     & HIP 103039   & HIP 103039   &  M4     &  3200  &   5.71 &    4.7  $\pm$    2.0 &   2.61  $\pm$   0.18 &   1.05 &    9.9  $\pm$    4.0 &   1.00  $\pm$   0.07 &   2.20 &       \\
M045A     & GJ 229       & HD 42581     &  M0.5   &  3805  &   5.75 &   11.7  $\pm$    2.5 &  11.68  $\pm$   0.12 &  -0.01 &    2.7  $\pm$    4.9 &   4.52  $\pm$   0.05 &  -0.37 &       \\
M046A     & GJ 693       & HIP 86990    &  M2     &  3267  &   5.84 &    2.5  $\pm$    2.0 &   3.05  $\pm$   0.07 &  -0.27 &   -4.1  $\pm$    5.1 &   1.19  $\pm$   0.03 &  -1.04 &       \\
M049A     & GJ 754       &              &  M4.5   &  3073  &   5.92 &    4.0  $\pm$    2.1 &   1.59  $\pm$   0.03 &   1.16 &    3.0  $\pm$    3.2 &   0.64  $\pm$   0.01 &   0.71 &       \\
M050A     & GJ 588       & HIP 76074    &  M2.5   &  3586  &   5.93 &    6.5  $\pm$    2.5 &   7.22  $\pm$   0.10 &  -0.29 &    5.4  $\pm$    6.3 &   2.78  $\pm$   0.04 &   0.42 &       \\
M051AB    & GJ 1005AB    & HIP 1242     &  M4     &  3153  &   5.95 &   -0.1  $\pm$    1.7 &   2.32  $\pm$   0.06 &  -1.43 &   -6.4  $\pm$    4.1 &   0.92  $\pm$   0.02 &  -1.77 &       \\
M052A     & GJ 908       & HIP 117473   &  M1     &  3706  &   5.95 &    3.8  $\pm$    1.8 &   5.79  $\pm$   0.12 &  -1.10 &    2.0  $\pm$    5.7 &   2.24  $\pm$   0.04 &  -0.05 &       \\
M053A     & GJ 285       & HIP 37766    &  M4.5   &  3235  &   5.98 &    6.7  $\pm$    2.1 &   3.81  $\pm$   0.04 &   1.37 &   13.2  $\pm$    3.9 &   1.46  $\pm$   0.02 &   2.97 &       \\
M054A     & GJ 268       & HIP 34603    &  M4.5   &  3200  &   6.12 &    5.8  $\pm$    2.0 &   3.74  $\pm$   0.45 &   1.02 &    8.4  $\pm$    5.7 &   1.44  $\pm$   0.17 &   1.21 &       \\
M055A     & GJ 555       & HIP 71253    &  M4     &  3185  &   6.22 &    4.6  $\pm$    1.7 &   3.31  $\pm$   0.21 &   0.71 &   11.4  $\pm$    4.7 &   1.27  $\pm$   0.08 &   2.18 &       \\
{\bf M056A} & GJ 581     & HIP 74995    &  M3     &  3374  &   6.34 &   21.1  $\pm$    1.5 &   3.29  $\pm$   0.15 & {\bf 12.12} & 22.1 $\pm$  5.0 &   1.26  $\pm$   0.06 & {\bf 4.19} & $<24$ \\
M057A     & GJ 896A      & HIP 116132   &  M3.5   &  3218  &   6.26 &    8.2  $\pm$    2.1 &   5.11  $\pm$   0.79 &   1.38 &   -2.2  $\pm$    5.2 &   1.98  $\pm$   0.31 &  -0.79 &       \\
M057B     & GJ 896B      & HIP 116132   &  M4     &  2872  &   6.26 &   -1.9  $\pm$    2.0 &   2.35  $\pm$   1.32 &  -1.74 &   -0.3  $\pm$    5.3 &   0.94  $\pm$   0.53 &  -0.23 &       \\
M058A     & LHS 2090     &              &  M6     &  2854  &   6.37 &   -0.1  $\pm$    2.1 &   0.41  $\pm$   0.04 &  -0.22 &   12.0  $\pm$    4.7 &   0.16  $\pm$   0.02 &   2.54 &       \\
M059A     & GJ 3737      &              &  M4.5   &  3263  &   6.38 &    2.4  $\pm$    2.1 &   0.90  $\pm$   0.07 &   0.71 &   -8.4  $\pm$    3.7 &   0.36  $\pm$   0.03 &  -2.35 &       \\
M060AB    & GJ 661A      & HD 155876    &  M3.5   &  3524  &   6.40 &    6.2  $\pm$    0.9 &   7.33  $\pm$   0.10 &  -1.28 &   -0.4  $\pm$    1.8 &   2.82  $\pm$   0.04 &  -1.82 &       \\
M061A     & GJ 3959      &              &  M5     &  2882  &   6.41 &    3.3  $\pm$    1.8 &   0.36  $\pm$   0.01 &   1.68 &    2.6  $\pm$    4.3 &   0.15  $\pm$   0.00 &   0.56 &       \\
M063A     & GJ 625       & HIP 80459    &  M1.5   &  3412  &   6.53 &    4.6  $\pm$    2.0 &   3.46  $\pm$   0.07 &   0.57 &    8.9  $\pm$    3.3 &   1.34  $\pm$   0.03 &   2.27 &       \\
M065A     & GJ 1156      &              &  M5     &  2962  &   6.54 &    2.5  $\pm$    2.0 &   0.87  $\pm$   0.01 &   0.81 &   -3.3  $\pm$    5.3 &   0.33  $\pm$   0.00 &  -0.68 &       \\
M066A     & GJ 3877      &              &  M7     &  2669  &   6.56 &   -3.1  $\pm$    1.7 &   0.26  $\pm$   0.01 &  -2.04 &   -6.7  $\pm$    3.8 &   0.12  $\pm$   0.00 &  -1.82 &       \\
M067A     & GJ 408       & HIP 53767    &  M2.5   &  3667  &   6.70 &    5.2  $\pm$    1.8 &   3.69  $\pm$   0.04 &   0.85 &    6.2  $\pm$    3.0 &   1.42  $\pm$   0.02 &   1.56 &       \\
M068A     & GJ 829       & HIP 106106   &  M3.5   &  3377  &   6.71 &    4.7  $\pm$    2.1 &   4.58  $\pm$   0.06 &   0.06 &    8.4  $\pm$    5.1 &   1.76  $\pm$   0.02 &   1.29 &       \\
M069A     & GJ 3522      &              &  M3.5   &  3128  &   6.77 &    3.0  $\pm$    2.0 &   4.47  $\pm$   0.08 &  -0.75 &    2.4  $\pm$    3.6 &   1.78  $\pm$   0.03 &   0.16 &       \\
M070A     & GJ 402       & HIP 53020    &  M4     &  3217  &   6.79 &    0.3  $\pm$    2.0 &   2.10  $\pm$   0.09 &  -0.93 &    6.0  $\pm$    3.5 &   0.81  $\pm$   0.03 &   1.46 &      \\
M071A     & GJ 880       & HD 216899    &  M1.5   &  3699  &   6.83 &    6.6  $\pm$    2.2 &   8.52  $\pm$   0.10 &  -0.87 &   -2.1  $\pm$    4.2 &   3.28  $\pm$   0.04 &  -1.30 &      \\
M072A     & GJ 299       &              &  M4.5   &  3088  &   6.84 &    5.2  $\pm$    1.8 &   0.76  $\pm$   0.01 &   2.50 &    8.1  $\pm$    3.3 &   0.31  $\pm$   0.01 &   2.38 &      \\
M073AC    & GJ 3192A     & HIP 14101    &  M2.5   &  3380  &   6.91 &    3.2  $\pm$    3.2 &   1.93  $\pm$   1.57 &   0.34 &    2.6  $\pm$    5.2 &   0.75  $\pm$   0.61 &   0.35 &      \\
M073B     & GJ 3192A     & HIP 14101    &  -      &  3470  &   6.91 &    2.8  $\pm$    2.8 &   2.11  $\pm$   0.03 &   0.25 &   -0.5  $\pm$    5.2 &   0.81  $\pm$   0.01 &  -0.24 &      \\
M074A     & GJ 1068      &              &  M4.5   &  3139  &   6.97 &    0.3  $\pm$    2.5 &   0.60  $\pm$   0.01 &  -0.10 &   -2.9  $\pm$    5.8 &   0.23  $\pm$   0.00 &  -0.54 &      \\
M076A     & GJ 393       & HIP 51317    &  M2     &  3574  &   7.13 &    5.2  $\pm$    2.2 &   4.51  $\pm$   0.05 &   0.30 &    2.9  $\pm$    4.8 &   1.74  $\pm$   0.02 &   0.24 &      \\
M077A     & GJ 4063      &              &  M3.5   &  3451  &   7.20 &    6.8  $\pm$    2.3 &   2.07  $\pm$   0.11 &   2.05 &    9.1  $\pm$    6.6 &   0.80  $\pm$   0.04 &   1.27 &      \\
M078A     & GJ 1286      &              &  M5.5   &  2810  &   7.23 &    4.1  $\pm$    2.1 &   0.49  $\pm$   0.01 &   1.70 &   10.5  $\pm$    3.8 &   0.21  $\pm$   0.00 &   2.70 &      \\
M079A     & GJ 4053      &              &  M4.5   &  3003  &   7.27 &    1.4  $\pm$    2.0 &   0.60  $\pm$   0.01 &   0.42 &    5.8  $\pm$    4.0 &   0.25  $\pm$   0.00 &   1.40 &      \\
{\bf M080A} & NLTT 54872 & LP 876-10    &  M4     &  3200  &   7.41 &    7.6  $\pm$    2.0 &   1.09  $\pm$   0.06 & {\bf 3.24} & 15.5 $\pm$   2.1 &   0.42  $\pm$   0.02 & {\bf 7.18} &  $<24$  \\
M081A     & GJ 4274      &              &  M4.5   &  2922  &   7.44 &    5.3  $\pm$    2.3 &   1.07  $\pm$   0.02 &   1.79 &    9.8  $\pm$    4.0 &   0.45  $\pm$   0.01 &   2.33 & $<24$ \\
M082A     & GJ 4248      &              &  M3.5   &  3190  &   7.45 &    3.8  $\pm$    2.1 &   1.64  $\pm$   0.04 &   1.02 &   -0.6  $\pm$    5.3 &   0.64  $\pm$   0.01 &  -0.24 &      \\
M083A     & GJ 3991      & HIP 83945    &  M3.5   &  3131  &   7.47 &   -2.4  $\pm$    2.4 &   2.13  $\pm$   0.04 &  -1.88 &   -5.2  $\pm$    4.8 &   0.82  $\pm$   0.01 &  -1.25 &      \\
M085A     & GJ 3378      &              &  M3.5   &  3206  &   7.57 &    4.5  $\pm$    1.8 &   1.76  $\pm$   0.03 &   1.50 &    4.9  $\pm$    4.9 &   0.69  $\pm$   0.01 &   0.86 &      \\
M087A     & GJ 514       & HIP 65859    &  M0.5   &  3777  &   7.59 &    7.3  $\pm$    1.8 &   5.31  $\pm$   0.06 &   1.12 &   -5.1  $\pm$    4.7 &   2.05  $\pm$   0.02 &  -1.51 &      \\
M088A     & GJ 3207      &              &  M3.5   &  3483  &   7.70 &   -2.3  $\pm$    1.5 &   0.18  $\pm$   0.01 &  -1.66 &   -1.3  $\pm$    4.5 &   0.07  $\pm$   0.00 &  -0.31 &      \\
M089A     & GJ 2005      &              &  M5.5   &  2549  &   7.71 &   -3.0  $\pm$    1.8 &   0.70  $\pm$   0.07 &  -2.05 &    5.8  $\pm$    4.0 &   0.27  $\pm$   0.03 &   1.38 &      \\
M090A     & GJ 1093      &              &  M5     &  2790  &   7.76 &    2.8  $\pm$    1.6 &   0.46  $\pm$   0.01 &   1.44 &   11.7  $\pm$    6.3 &   0.20  $\pm$   0.00 &   1.83 & $<24$ \\
M092A     & GJ 480.1     & HIP 61874    &  M3.0   &  3224  &   7.78 &    5.0  $\pm$    2.3 &   0.89  $\pm$   0.03 &   1.78 &    2.4  $\pm$    3.3 &   0.35  $\pm$   0.01 &   0.64 &      \\
M093A     & GJ 54        & HIP 5496     &  M2     &  3479  &   7.80 &    6.7  $\pm$    1.8 &   5.61  $\pm$   0.07 &   0.59 &    2.2  $\pm$    2.9 &   2.15  $\pm$   0.03 &   0.01 &      \\
M094AB    & GJ 831A      & HIP 106255   &  M4.5   &  3044  &   7.80 &    3.7  $\pm$    1.8 &   2.42  $\pm$   0.04 &   0.70 &   12.7  $\pm$    4.2 &   0.94  $\pm$   0.02 &   2.77 &      \\
M095A     & GJ 382       & HIP 49986    &  M1.5   &  3674  &   7.93 &    7.0  $\pm$    1.7 &   5.53  $\pm$   0.07 &   0.83 &    1.0  $\pm$    4.3 &   2.13  $\pm$   0.03 &  -0.26 &      \\
M096A     & GJ 300       &              &  M4     &  3113  &   7.96 &   -4.0  $\pm$    2.0 &   1.77  $\pm$   0.10 &  -2.82 &   -6.0  $\pm$    4.4 &   0.68  $\pm$   0.04 &  -1.52 &      \\
M098A     & GJ 623       & HIP 80346    &  M2.5   &  3435  &   8.03 &    3.9  $\pm$    1.6 &   3.14  $\pm$   0.13 &   0.49 &    3.8  $\pm$    4.0 &   1.21  $\pm$   0.05 &   0.64 &      \\
M099AB    & GJ 257A      & HIP 33499    &  M3     &  3310  &   8.03 &    2.0  $\pm$    2.4 &   2.84  $\pm$   0.12 &  -0.34 &   -2.4  $\pm$    5.8 &   1.09  $\pm$   0.05 &  -0.60 &      \\
M100A     & GJ 686       & HIP 86287    &  M1     &  3629  &   8.05 &    2.3  $\pm$    2.1 &   3.77  $\pm$   0.12 &  -0.69 &   11.6  $\pm$    4.9 &   1.45  $\pm$   0.04 &   2.07 &      \\
M101A     & GJ 1289      &              &  M4     &  3109  &   8.10 &    1.1  $\pm$    2.1 &   1.08  $\pm$   0.02 &  -0.01 &   -4.6  $\pm$    6.2 &   0.43  $\pm$   0.01 &  -0.81 &     \\
M103A     & GJ 493.1     &              &  M4.5   &  3020  &   8.12 &   -1.4  $\pm$    1.9 &   0.75  $\pm$   0.01 &  -1.10 &  -11.8  $\pm$    4.4 &   0.30  $\pm$   0.01 &  -2.77 &     \\
M104AB    & GJ 747A      &              &  M3     &  3413  &   8.18 &   -0.3  $\pm$    1.6 &   2.00  $\pm$   0.12 &  -1.40 &   -7.5  $\pm$    3.8 &   0.77  $\pm$   0.05 &  -2.19 &     \\
M106A     & GJ 1151      &              &  M4.5   &  2996  &   8.19 &   -1.0  $\pm$    2.2 &   0.78  $\pm$   0.02 &  -0.79 &   -3.9  $\pm$    5.4 &   0.33  $\pm$   0.01 &  -0.79 &     \\
M107A     & GJ 1227      &              &  M4.5   &  3006  &   8.23 &   -2.2  $\pm$    2.0 &   0.70  $\pm$   0.01 &  -1.45 &    8.0  $\pm$    4.1 &   0.29  $\pm$   0.01 &   1.87 &     \\
M109A     & GJ 1105      & HIP 38956    &  M3.5   &  3176  &   8.27 &   -0.9  $\pm$    1.8 &   1.44  $\pm$   0.03 &  -1.31 &    4.9  $\pm$    4.2 &   0.56  $\pm$   0.01 &   1.05 &     \\
M110AB    & GJ 1230A     &              &  M5     &  3150  &   8.27 &    3.1  $\pm$    1.6 &   2.17  $\pm$   0.06 &   0.56 &   -1.9  $\pm$    3.6 &   0.84  $\pm$   0.02 &  -0.76 &     \\
M112A     & GJ 486       & HIP 62452    &  M3.5   &  3222  &   8.36 &    5.0  $\pm$    2.0 &   2.24  $\pm$   0.03 &   1.36 &    3.2  $\pm$    4.2 &   0.87  $\pm$   0.01 &   0.55 &     \\
M114A     & GJ 1154AB    &              &  M5     &  2882  &   8.38 &   -2.7  $\pm$    1.9 &   0.88  $\pm$   0.02 &  -1.87 &    5.8  $\pm$    3.7 &   0.37  $\pm$   0.01 &   1.47 &     \\
M115A     & GJ 3146      &              &  M5.5   &  2732  &   8.50 &   -4.0  $\pm$    1.6 &   0.23  $\pm$   0.00 &  -2.59 &   -1.0  $\pm$    3.7 &   0.10  $\pm$   0.00 &  -0.29 &     \\
M116A     & GJ 1057      &              &  M5     &  2927  &   8.54 &    2.1  $\pm$    2.3 &   0.64  $\pm$   0.01 &   0.61 &    4.2  $\pm$    4.5 &   0.27  $\pm$   0.00 &   0.87 &     \\
M117A     & GJ 3454      &              &  M5     &  2974  &   8.58 &    0.9  $\pm$    1.5 &   1.12  $\pm$   0.04 &  -0.12 &    5.5  $\pm$    2.8 &   0.43  $\pm$   0.01 &   1.81 &     \\
\hline
\end{longtable}
\end{landscape}

\end{document}